\documentclass[12pt]{article}
\pdfoutput=1 
\usepackage{setspace}
\usepackage[T1]{fontenc}
\usepackage{lmodern}
\usepackage[colorlinks=true, pdfstartview=FitV, linkcolor=blue,citecolor=blue, urlcolor=blue]{hyperref}

\usepackage{microtype}
\usepackage[margin=1in]{geometry}
\usepackage{setspace}
\usepackage[title]{appendix}
\usepackage[round]{natbib}
\usepackage[dvipsnames]{xcolor}
\usepackage{amsmath,amsthm,amssymb,ifthen,mathrsfs,amsfonts,bm}
\usepackage{empheq}
\usepackage{mathtools}
\usepackage{subcaption}
\usepackage{graphicx}
\usepackage{enumitem}
\usepackage{bigints}
\usepackage{bookmark}
\usepackage{chngcntr}
\usepackage{tocloft}

\usepackage[framemethod=tikz]{mdframed}
\usepackage{pgf,tikz}
\usetikzlibrary{calc,math}
\usetikzlibrary{arrows,shapes,positioning,automata}
\usetikzlibrary{decorations.markings}
\usetikzlibrary{shapes.geometric}
\usepackage{pgfplots}
\usetikzlibrary{intersections}
\pgfplotsset{compat=newest}
\usepgfplotslibrary{fillbetween}
\usetikzlibrary{patterns}

\newtheorem{theorem}{Theorem}
\newtheorem{lemma}{Lemma}
\newtheorem{proposition}{Proposition}
\theoremstyle{definition}
\newtheorem{corollary}{Corollary}
\newtheorem{definition}{Definition}

\newtheorem{notation}{Notation}

\newtheorem{example}{Example}

\newtheorem{remark}{Remark}

\def\BE{\mathsf{E}}
\def\BP{\mathsf{P}}
\def\BR{\mathbb{R}}

\def\d{\mathrm{d}}
\def\ee{\mathrm{e}}
\def\ve{\varepsilon}

\newcommand{\df}[1]{\textit{#1}}

\numberwithin{equation}{section}

\makeatletter
\newcommand{\rn}[1]{\romannumeral #1}
\newcommand{\Rn}[1]{\expandafter\@slowromancap\romannumeral #1@}
\makeatother

\author{Farzad Pourbabaee and Federico Echenique}
\title{Binary Mechanisms under Privacy-Preserving Noise\thanks{Pourbabaee (\href{mailto:far@caltech.edu}{far@caltech.edu}) is at the Division of the Humanities and Social Sciences, Caltech. Echenique (\href{mailto:fede@econ.berkeley.edu}{fede@econ.berkeley.edu}) is at the Department of Economics, UC Berkeley.}}
\date{May 2024}
\renewcommand\footnotemark{}
\begin{document}
\maketitle
\begin{abstract}
        We study mechanism design for public-good provision under a noisy privacy-preserving transformation of individual agents' reported preferences. The setting is a standard binary model with transfers and quasi-linear utility. Agents report their preferences for the public good, which are randomly  ``flipped,'' so that any individual report may be explained away as the outcome of noise. We study the tradeoffs between preserving the public decisions made in the presence of noise (noise sensitivity), pursuing efficiency, and mitigating the effect of noise on revenue. 
    \end{abstract}
\clearpage
\tableofcontents
\newpage

\section{Introduction}
The field of mechanism design considers agents who hold private information about their preferences. Agents are asked to surrender this information when properly incentivized; but, traditionally, mechanism design ignores any potential {\em privacy concerns} that may add to agents' reluctance to reveal their true preferences. 

Privacy concerns may arise because agents have an intrinsic ``non-instrumental'' aversion to revealing their preferences, or because agents worry that such information can be used against them in future interactions. This may occur in the provision of both private and public goods. With private goods, an agent who surrenders their willingness to pay for a good to a seller will lose surplus in future interactions with this seller. 

In public-goods settings, revealing willingness-to-pay today may imply higher future taxes for related public goods tomorrow. For example, revealing a high value for a playground today may reveal a high value for a public library in the future. Agents may also have non-instrumental preferences for privacy:  public-health-related projects often involve sensitive information about agents' likelihood of being susceptible to disease (think, for example, of a cancer screening program). In such cases, we want to know when we can preserve individuals' privacy while minimally compromising the optimality of a public choice rule.

Our paper studies a specific mechanism design problem: a planner faces a standard binary public-good provision problem with quasilinear utility and monetary transfers. Our planner cares about individuals' preferences for the public good, and about the revenue they can collect from its provision (alternatively, the extent to which they need to subsidize the public good). In our version of the problem, privacy concerns are important, and dealt with using an embedded privacy-preserving operation.

There are a variety of proposals to protect individuals' privacy. Our approach in this paper involves adding random noise to the individuals' messages, using an \textit{in-place} randomization device, before they reach the planner. In effect,  the planner's mechanism takes as inputs the agents' perturbed reports --- so it can only access the agents' reported types after they have been subject to a privacy-preserving random transformation. The idea follows the literature on {\em differential privacy} initiated by \citet*{dwork2006}, \citep[see also][]{dwork2006data,dwork2008differential,dworkroth2014}, by which individuals' privacy is preserved through the addition of random noise. 

There are other privacy-preserving methods as well. For example, the method of de-identification, which involves removing names and personal identifying information from messages. However, as shown in a recent study by~\cite{evans2023statistically}, de-identification in the context of political surveys is vulnerable to re-identification attacks, compromising the privacy of survey respondents.\footnote{De-identification has been ineffective in other areas as well. For example, in a study by~\cite{gymrek2013identifying}, researchers successfully re-identified individuals from the 1000 Genomes Project by cross-referencing their data with other publicly available online resources.}

Other methods of privacy protection require the availability of a trusted intermediary. This intermediary may use cryptographic tools. In this approach, agents report their types to the intermediary, who aggregates the information and passes along a privacy-preserving aggregate decision to the planner. The intermediary may be automated, and consist of a cryptographic algorithm, but still needs to be trusted by the agents involved. The downside of this method is that assuming the existence of a trusted intermediary can be quite restrictive. Additionally, such intermediaries may act strategically, and extract rents from the agents. For example, when agents reveal willingness to pay for a public good, a trusted and benevolent principal with imperfect commitment ability may extract future rent from some individual agents.

Aside from evaluating the pros and cons of other methods, our analysis in this paper focuses on the positive aspects. We believe that understanding the tradeoffs in a noisy environment is a valuable exercise in its own right.

In our paper, agents' types are binary and encode how much utility they receive from the public good: high or low. Types take the value $+1$ or $-1$. The provision outcome is also binary, i.e, $\{0,1\}$-valued. Binary decisions are common in public goods environments because public goods are often about implementing a large indivisible project (a library, a bridge, a waste disposal facility, etc.). We focus on this binary setting, because the basic tradeoffs are captured by a yes/no decision. Therefore our model gets at the heart of the matter, while remaining tractable.

The random noise is then simply a ``flip,'' which occurs with probability $\delta\in (0,1/2)$. If an agent reports a type $x_i\in\{-1,+1\}$, then the mechanism receives $x_i$ with probability $1-\delta$, and a ``flipped'' report $-x_i$ with probability $\delta$. As a consequence, an agent can always explain away any evidence about their type as the outcome of a random flip. Their explanation is more credible the larger the value of $\delta$. Noise is then desirable because larger values of $\delta$ offer a better protection of privacy.

The problem with adding noise to the agents' reports---one might say the flip side---is, of course, that the quality of the planner's decision suffers. So we consider the probability that the planner's decision is affected by the noise we have added for reasons of privacy. A key concept is {\em noise sensitivity:} the probability that the planner's decision differs from what it would have been, given the true and noise-free reports. In addition to standard considerations in mechanism design (such as revenue and surplus), our paper evaluates mechanisms on the basis of their noise sensitivity. 
We further show that noise sensitivity is closely tied to the distortions in social surplus caused by random transformations of the agents' reports.

Noise affects transfers, as well as the public-good provision decision. In consequence, the planner's expected revenue suffers. Standard ideas in mechanism design mean that an agent with a low value for the public good (a $-1$ type) pays less than an agent with a high value (a $+1$ type). With noise, it is possible that a low type has their truthful report flipped, and is thus subject to the higher payment designed for high types. This, in turn, affects the whole problem by means of the low types' participation constraints. The end result is lower revenue for the planner as a whole. In sum, as the level of noise $\delta$ increases, revenue and social surplus decrease. There is thus a tradeoff between the privacy protection afforded by noise, and the effectiveness of a given social choice function in terms of traditional economic objectives. 

Our first main theorem concerns choosing a mechanism to minimize noise sensitivity (or equivalently, surplus distortion), given a target level of revenue and a fixed level of privacy-preserving noise. The resulting optimization is not convex, which presents a challenge, but we are able to characterize the optimal mechanisms asymptotically, as the number of agents grows. They take the form of linear threshold functions (basically implementing the public good once the number of ``votes'' in favor, or high types, exceeds those against by a certain margin). In our second main theorem, we characterize the mechanism that optimizes social surplus under a target revenue constraint, and a fixed noise level. The optimal provision rule is a linear threshold function and coincides with one of the optimal solutions from the earlier problem. We find the provision threshold as a function of the required revenue and the noise level. Moreover, it is shown that the lower the provision threshold, the higher the social surplus, and the smaller the revenue is. Together, these two theorems examine the \textit{level} and \textit{distortions} of social surplus under a revenue constraint across varying levels of privacy-preserving noise.

A key tradeoff in our paper involves noise sensitivity and revenue. A planner can make the mechanism more robust to noise (improve its noise sensitivity) at the cost of lower revenue. In sum, our paper describes a planner who balances several different objectives: privacy, efficiency, robustness and revenue.\footnote{We use the revenue terminology throughout the paper, but one may of course think of the objective as minimizing the amount of subsidy needed for the public good project.} The different tradeoffs involved are characterized through the theory developed in the paper.

Our model has two other interpretations, in addition to the emphasis on privacy that we have focused on so far. First, agents may be unable to perfectly communicate their preferences to the mechanism. Miscommunication has been documented experimentally \citep[see][]{budish2022can}, and pushed as an agenda by, for example, \cite{mcfadden2009}. Second, agents may have imperfect information about their own preferences. Their reports are therefore only noisy versions of their underlying values for the public good.  Imperfect knowledge of preferences has been considered a key motivation for studying information acquisition in mechanism design. Two recent examples are \cite{gleyze2023} and \cite{thereze2023}. To summarize, the one formal framework that we introduce and study provides insights about three important environments.


\paragraph{Related Literature.}
We are not the first to study mechanism design together with a device for ensuring privacy. There is a literature on mechanism design and differential privacy. The first paper is \cite{mcsherry2007mechanism}, who shows that differential privacy can be a useful tool in obtaining incentive compatibility. By dampening the effect that any individual report has on the mechanism's decision, differential privacy can help ensure truthful behavior among agents. \cite{nissim2012approximately} develop these ideas in a construction that achieves approximately optimal virtual implementation. Their focus is therefore closer to the problem of full implementation, and not the standard mechanism design problem. \cite{huang2012exponential} propose mechanisms that are both incentive compatible and differentially private, but does not incorporate the analysis of the tradeoffs that are the focus of our paper. The works of  \cite{nissim2012approximately}, \cite{xiao2013privacy}, and \cite{chen2016truthful} all consider preferences over privacy explicitly in their mechanism design analysis. This is of course an important direction, but not the one we pursue here. \cite{nissim536761} provide an overview of the literature on mechanism design and differential privacy. 

Our paper is also related to recent works on monopolistic screening with privacy concerns \citep{eilat2021bayesian, krahmer2023optimal}. In the first paper, the privacy loss --- measured by the Kullback-Leibler divergence between planner's prior and posterior belief about the buyer's type --- is set as a constraint for the screening problem. Specifically, in this work the privacy is protected by selecting the message space as the partitions of the original type space (i.e., coarsening the type set). Hence, the message sent by the agent does not fully resolve the underlying type, thus protecting their privacy. In our binary setting, noisy flips are more natural than a partition of the type space, which is too blunt when there are only two types. The second paper reflects privacy concerns in the buyer's preference, much like the literature we discussed above. Neither of the papers address the tradeoffs that we focus on, or the issues regarding robustness. 

The idea of adding noise as a means for privacy protection is very common in other areas as well \citep[e.g., see][for applications in communication and information theory]{geng2015optimal, he2018preserving}. In political science \cite{warner1965randomized} introduced the randomized response method as a survey technique, that asks respondents to use in-place randomization device to conceal their sensitive answers from the interviewer --- \cite{blair2015design} summarizes the use of this method in this area. Since other methods of privacy protection (such as clean rooms and de-identification) have been shown to fail, differential privacy through the addition of calibrated noise gained traction in political science. In a sequence of studies by \cite{evans2019statistically}, \cite{evans2022differentially} and  \cite{evans2023statistically} this method is shown to help social scientist to study the vast amount of user data owned by governments and companies while maintaining privacy issues. For example, the last US Census issued by the government is being released with noise.\footnote{See \href{https://www2.census.gov/about/policies/2019-11-paper-differential-privacy.pdf}{https://www2.census.gov/about/policies/2019-11-paper-differential-privacy.pdf}.} Companies also use open source softwares that allow researchers to test their algorithms while concealing the private data of their users through the addition of statistical noise.\footnote{See \href{https://news.microsoft.com/on-the-issues/2020/08/27/statistical-noise-data-differential-privacy}{https://news.microsoft.com/on-the-issues/2020/08/27/statistical-noise-data-differential-privacy}.}

Our model of public good provision with privacy-protection concerns is also formally equivalent to a setting in which agents cannot perfectly report their preferences to the planner. In that sense our paper is a theoretical contribution to a mostly empirical literature that documents preference \textit{misrepresentation} in incentive compatible environments because of variety of reasons such as cognitive limitations or simply lack of perfect communication between participants and the planner. In his tribute to Hurwicz and Laffont, \cite{mcfadden2009} states that ``in reality, mistakes that agents make in processing and
drawing inferences from communications and information, and in exercising control
and responding to incentives, can undermine the ideal efficiency of mechanisms, making
it important to consider the robustness of mechanisms involving human agents.''

A growing body of literature in applied mechanism design documents preference misrepresentation. For example, \cite{hassidim2017mechanism} and \cite{hassidim2021limits} show that students misreport their funding preferences when applying to graduate programs, despite the fact that the underlying matching mechanism is strategy-proof (in this case it is Deferred-Acceptance). In the context of residency matching mechanisms \cite{rees2018suboptimal} and \cite{rees2018experimental} present evidences that some students make futile attempts misrepresenting their preference ranking. In an experiment \cite{budish2022can} show that students fail to report their preferences accurately enough in a course scheduling mechanism.

\section{Model}
\label{sec:model}
\subsection{Mechanisms}
We consider the problem of providing a public good in an economy with $n$ agents and
quasilinear preferences. The decision is binary: a public good is either provided or
not. Agents' types, which are denoted by $x_i\in\{-1,+1\}$, encode their value for
the public good. An individual with a low (respectively, high) type has low
(respectively, high) valuation for the public good. Ideally, a decision on whether to
provide the public good is based on agents' realized types, but these are private
information. We have access to monetary transfers that may be used to incentivize
agents in reporting their types. The assumption of binary types is, of course,
restrictive, but crucial for the methodology employed in our paper; it also offers
the simplest framework for studying a public choice decision with heterogeneous and
private preferences over the provision of the public good. 

We focus on direct-revelation mechanisms.  A (direct-revelation) public-good mechanism consists of an \textit{allocation rule} $f:\{-1,+1\}^n \to \{0,1\}$, and $n$ \textit{transfer rules}, denoted by $t_i:\{-1,+1\}^n \to \BR$ for all $i \in [n]$. The allocation rule $f$ takes in the $\{-1,+1\}$ messages sent by the individuals, and returns the provision decision, where an output of $1$ means the public good is being provided, and a $0$ output means otherwise. Often in the paper we call an allocation rule a \textit{social choice function} (SCF).

\subsection{Quasilinear Preferences}

A profile of types $(x_1,\ldots,x_n)$ is drawn  i.i.d.\ from the uniform distribution on $\{-1,+1\}$.\footnote{The measure does not need to be uniform. In fact, it is possible to change the type domain to any other bi-valued set with un-even probability ––– that just requires some scaling and normalization. For example, if the type space is $\{\ell, h\}$ with probabilities $\{p, 1-p\}$, we can can transform the preference bias from $b$ to $b'$ and the range of $f$ from $\{0,1\}$ to $\{0, M\}$ so that transformed problem becomes isomorphic to the original problem. We chose the convention of the uniform measure over $\{-1,+1\}$ because it is standard in the Boolean function literature.} Individuals have quasilinear preferences over the final allocation and the transfer. Specifically, the utility of individual $i$,  with type  $x_i$, from $(f,t_i)\in\{0,1 \}\times\BR$ is
\begin{equation}
\label{eq:utility_form}
    u_i(f,t_i;x_i) = \left(\frac{b+x_i}{2}\right) f-t_i\,.
\end{equation}
The parameter $b\in [0,1]$ captures a possible bias in favor of the public good. For
example, when $b=1$, the efficient outcome is to always provide the public good, and
when $b=0$, the preferences for the public good are \textit{symmetric} around zero, and the efficient outcome coincides with a simple majority decision. The negative sign before $t_i$ means that the transfers are from the individuals to the planner. 
\subsection{Noisy Reports}
A key innovation in our paper is the presence of noisy preference reports. Specifically, we assume the message $m_i \in \{-1,+1\}$ sent by individual $i$ is going to flip to $-m_i$ with probability $\delta \in (0,1/2)$. We assume these flips are independent across all individuals, and refer to $\delta$ as the \df{noise probability}. Agents can explain away any information about their type as the result of these random flips. Such explanations are more credible the larger the value of $\delta$.  Noise in our model is a basic implementation of differential privacy \citep{dwork2006,dworkroth2014}. When $\delta$ is close to $1/2$, each individual agent's report is approximately uniformly distributed on $\{-1,+1\}$, regardless of their actual report.

A basic inspiration for differential privacy is the model of ``randomized response'' used in survey studies in the social sciences, see Chapter 2 in \cite{dworkroth2014}. In our setting, let $m_i$ be the message sent by agent $i$, and $y_i$ be the signal received by the planner. Then, following the language of differential privacy, this communication mechanism will be \textit{$\ve$-differentially private} if
\begin{equation}
\label{eq:epsilon_diff}
    \ee^{\ve} \geq \max\left\{\frac{\BP\left(y_i= +1 | m_i = +1\right)}{\BP\left(y_i = +1 | m_i = -1\right)}\, , \,  \frac{\BP\left(y_i= -1 | m_i = +1\right)}{\BP\left(y_i = -1 | m_i = -1\right)}\right\} = \frac{1-\delta}{\delta}\,.
\end{equation}
It essentially means that by observing the received signal, the planner cannot distinguish the transmitted message with high precision. Thus by setting $\delta = (1+\ee^{\ve})^{-1}$ our model guarantees an $\ve$-differentially private mechanism.

An alternative explanation for noisy preference reporting is that communication from the agents to the social planner can be lossy and imperfect; hence, random flips capture imperfect communication between the agents and the planner. If the privacy interpretation of our model makes sense when $\delta$ is large, the lossy communication interpretation makes most sense when $\delta$ is small. 

\subsection{Quantifications}
\label{subs:quant}
In this section, we briefly discuss the implementability notions and the main quantities that we use in the paper. Subsequently, in the following sections, we delve deeper into their formal definitions

The first quantity that we introduce is \textit{noise sensitivity}. In a Bayes-Nash
incentive compatible (BN-IC) mechanism, each agent reports their true type $x_i \in
\{-1, +1\}$, but as a result of noisy preference reporting, the planner receives $y_i
\in \{-1, +1\}$, where $\BP(x_i \neq y_i)=\delta$. Observe that the pairs $(x_i,
y_i)$ are i.i.d.\ over $i\in [n]$. In particular, $x = (x_1, \ldots, x_n)\sim
\text{Unif}\big(\{-1,+1\}^n\big)$ and $y=(y_1, \ldots,y_n)$ is the noisy version of
$x$ received by the planner. This means that the implemented outcome that was
supposed to be $f(x)$, now changes to $f(y)$. If $f(x)$ is the desired decision
regarding the public good, we may be concerned that $f(y)\neq f(x)$. The probability
that this occurs is termed the  noise sensitivity of the SCF $f$.

Specifically, \df{noise sensitivity} is defined as
\begin{equation*}
    \mathsf{NS}_\delta[f] = \BP\big(f(x) \neq f(y)\big)\,.
\end{equation*}
Noise sensitivity is a standard variable in the analysis of Boolean functions (see, e.g., the pioneering work by~\cite{benjamini1999noise} and the comprehensive treatment in~\cite{o2014analysis}). We believe that this quantity is important in and of itself. For one, the planner does not want to pick an allocation rule that frequently takes the individuals by surprise. This would affect the credibility and commitment power of the planner. 

Second, in Section~\ref{sec:noise_in_alloc}, we show that as the size of the
economy grows, the distortions in social surplus caused by differential privacy noise
can be closely approximated by the noise sensitivity (see, in particular, Proposition~\ref{prop:surplus_dist}). Third, increasing the noise level
$\delta$ adds to the privacy preservation power of the mechanism, at the expense of
making the SCF more sensitive to the noise. Studying the dependence of noise
sensitivity on $\delta$ quantifies the tradeoff between privacy and the ensuing distortion.

Next, we present the other two quantities: social surplus and revenue. Suppose that, by refusing to participate in the mechanism, any individual can guarantee themselves a utility of zero. A mechanism that respects the interim individual rationality constraint is referred to by IIR. We say that a SCF $f$ is Bayes-Nash implementable if it is both BN-IC and IIR. For such a SCF we refer to its expected social surplus by $\mathsf{S}_\delta[f]$, and to the maximum expected revenue by $\mathsf{R}_\delta[f]$. The expectation and probability operators are with respect to the joint distribution of $x$ and $y$.

\section{Main Results}
\label{sec:results}
We introduced three main quantities in the previous section: expected social suplus
$\mathsf{S}_\delta[f]$, maximum expected revenue $\mathsf{R}_\delta[f]$, and noise
sensitivity $\mathsf{NS}_\delta[f]$. Understanding these quantities in a finite
economy is very challenging, but we shall see that the problem is tractable in a
large economy.  We now proceed with the two main results in the
paper, which examine the tradeoffs between these quantities as the number of individuals grows large (i.e., $n \to \infty$).
\subsection{Revenue and Noise Sensitivity}

The concerns for robustness in the presence of reporting noise motivates a natural
optimization problem. Among the set of all implementable allocation rules that extract a target level of expected revenue (say $R$), which ones have the minimum noise sensitivity (or maximum noise robustness)? Formally, we seek the solution to the following optimization problem:

\begin{equation}
\label{eq:optimization_prob}
\begin{gathered}
     \min_f \mathsf{NS}_\delta[f]\\
     \text{subject to: }\, \mathsf{R}_\delta[f] \geq R \text{ and } f \text{ being implementable}\,.
\end{gathered}
\end{equation}
The solution to problem~\eqref{eq:optimization_prob} characterizes the tradeoff between privacy and
expected revenue in public-good mechanisms. Specifically, raising the noise level
$\delta$ provides higher privacy, but increases the noise sensitivity, and (as will be
shown later) decreases revenue. Fixing the noise level $\delta$, thereby guaranteeing
a certain privacy preservation level, the above program outputs the SCF that raises
the target revenue $R$ and is maximally robust against the privacy-preserving noise induced through~$\delta$.

Obtaining a closed-form solution to Problem~\eqref{eq:optimization_prob} is not tractable, but we can make progress under the assumption of large $n$.

Before stating the solution to Problem~\eqref{eq:optimization_prob}, we state some notational conventions.
\begin{notation}
\label{not:gaussians}
We use $\varphi(\cdot)$ and $\Phi(\cdot)$ to respectively denote the density and cumulative distribution function of the standard Gaussian. Also, we denote the inverse function of the Gaussian density (taking values in $\BR_+$) by $\varphi^{-1}$, and the inverse function of the Gaussian cumulative function by $\Phi^{-1}$. We further denote the sum of individuals' types by $\nu_n(x) \coloneq  \sum_{i = 1}^n x_i$, and sometimes drop $x$ from the argument of $\nu_n(\cdot)$.
\end{notation}

\begin{notation}
\label{not:normalization}
We normalize the target revenue $R$, and define $r \coloneq R/(1-2\delta)\sqrt{n}$. The optimal value of the program in~\eqref{eq:optimization_prob} is denoted by $\mathcal{V}_n(r)$.
\end{notation}

\begin{definition}
A mapping from $\{-1, +1\}^n$ to $\{0,1\}$ is called a \textit{linear threshold function} (LTF), if there exists some threshold $\tau$, such that $f(x) = \mathbf{1}\left\{\nu_n(x) \geq \tau\right\}$. In the following, we mainly work with two LTFs:
\begin{subequations}
\begin{align}
    \label{eq:upper_LTF}
    \bar{\ell}_n(x;r) &\coloneq \mathbf{1}\left\{\frac{\nu_n(x)}{\sqrt{n}} \geq \varphi^{-1}(r)+o(1)\right\}\,,\\
    \label{eq:lower_LTF}
    \underline{\ell}_n(x;r) &\coloneq \mathbf{1}\left\{\frac{\nu_n(x)}{\sqrt{n}} \geq -\varphi^{-1}(r)+o(1)\right\}\,,
\end{align}
\end{subequations}
where, as usual, $o(1)$ denotes a term that vanishes as $n\to \infty$.
\end{definition}

\begin{theorem}
\label{thm:main}
The linear threshold functions $\{\underline{\ell}_n(\cdot;r),\bar{\ell}_n(\cdot;r)\}$ are asymptotically optimal choices for the revenue constrained noise sensitivity minimization problem in \eqref{eq:optimization_prob}. Formally, 
\begin{equation}
\label{eq:main}
    \mathcal{V}_n(r)\leq \mathsf{NS}_\delta[\ell_n] \leq \mathcal{V}_n(r)+o(1),\,\, \text{ for } \ell_n \in \{\underline{\ell}_n(\cdot;r),\bar{\ell}_n(\cdot;r)\}\,.
\end{equation}
\end{theorem}

In the following sections, we argue that the simple majority rule (a LTF with $0$ threshold, or $50\%$ of the votes) raises the maximum revenue, but if one wants to improve upon its noise sensitivity, then by Theorem~\ref{thm:main} the optimal way, among all implementable Boolean functions, is to increase the 50\% threshold of the majority function (or decrease it by a similar amount). The more one increases (or decreases) this threshold, the more noise robustness is gained and more expected revenue is lost. The optimal tradeoff is struck by the LTFs $\{\underline{\ell}_n(\cdot;r),\bar{\ell}_n(\cdot;r)\}$.

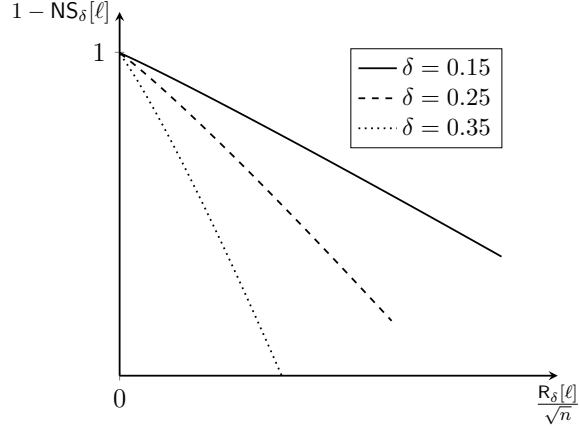
\begin{figure}[ht]
\begin{center}
\begin{tikzpicture}[scale=0.85]
    \begin{axis}[
    axis lines=left,
    smooth,
    axis line style={thick},
    xtick={0.001},
    xticklabels={$0$},
    ytick={1},
    yticklabels={$1$},
    xmax = 0.32,
    ymax = 1.05,
    x label style={at={(axis description cs:1,0)},anchor=north},
	xlabel={$\frac{\mathsf{R}_\delta[\ell]}{\sqrt{n}}$},
	y label style={at={(axis description cs:0,1)},anchor=east,rotate=-90},
	ylabel={\footnotesize{$1-\mathsf{NS}_\delta[\ell]$}},
	legend style={at={(axis description cs: 0.7,0.9)},anchor=north,font=\small},
    ]
    \addplot[thick ] table[col sep=comma] {NS_revenue_one.csv};
    
    \addplot[thick, dashed] table[col sep=comma] {NS_revenue_two.csv};
     
    \addplot[thick, dotted] table[col sep=comma] {NS_revenue_three.csv};
     
    \legend{$\delta=0.15$,$\delta=0.25$,$\delta=0.35$};
    \end{axis}
\end{tikzpicture}
\caption{Asymptotic Pareto Frontier}
\label{fig:pareto_frontier}
\end{center}
\end{figure}

In Figure \ref{fig:pareto_frontier}, we plot the maximum achievable noise robustness, i.e., $1-\mathsf{NS}_\delta$, on the $y$-axis, given the revenue level on the $x$-axis. We use the normalized expected revenue (by $\sqrt{n}$ not $(1-2\delta)\sqrt{n}$). The figure indicates the asymptotic Pareto frontier, for three different noise levels, that are achieved by the LTFs in Theorem~\ref{thm:main}. The figure suggests that, as the noise \textit{increases}, the frontier becomes \textit{steeper}; meaning that giving up a fixed level of revenue can lead to greater robustness against noise, and this tradeoff is amplified in higher noise levels where the privacy protection is stronger.

Put differently, the maximum achievable noise robustness is decreasing with respect to both revenue and level of noise. However, these two variables act as \textit{substitutes}. That is, lowering the required revenue is more effective for gaining noise robustness at higher levels of noise.
\subsection{Revenue and Surplus}
Our second main result deals with the tradeoff between revenue and surplus. Specifically, we ask and answer the following question: For a fixed level of privacy noise $\delta$, and among all the implementable SCFs that raise a target expected revenue (say $R$), which SCF has the highest social surplus? Formally, we study the following optimization problem:
\begin{equation}
\label{eq:surplus_rev_optimization}
\begin{gathered}
     \max_f \mathsf{S}_\delta[f]\\
     \text{subject to: }\, \mathsf{R}_\delta[f] \geq R, \text{ and } f \text{ being implementable}\,.
\end{gathered}
\end{equation}
\begin{theorem}\label{thm:surplus_rev}
For large enough $n$, the optimal solution in the revenue/surplus tradeoff in~\eqref{eq:surplus_rev_optimization} is the linear threshold function $\underline{\ell}_n(\cdot;r)$ expressed in~\eqref{eq:lower_LTF}.
\end{theorem}

In contrast with Theorem~\ref{thm:main}, Theorem~\ref{thm:surplus_rev} provides an allocation rule that is \textit{exactly} optimal when $n$ is large enough, whereas the candidates in Theorem~\ref{thm:main} are \textit{asymptotically} optimal, as stated in equation~\eqref{eq:main}.

Theorem~\ref{thm:surplus_rev} states that, if one is willing to sacrifice some revenue (compared to the maximum achievable under the majority rule), the optimal approach to maximize expected social surplus is to lower the majority threshold below $50\%$. The lower the provision threshold, the higher the expected social surplus, and the smaller the expected revenue. Additionally, this tradeoff is optimally struck by the threshold function $(R,\delta)\mapsto -\varphi^{-1}(R/(1-2\delta)\sqrt{n})$, which is the provision threshold in $\underline{\ell}_n(\cdot;r)$.

Importantly, fixing a target revenue level $R$, one observes that securing the mechanism by increasing the noise level $\delta$, raises the provision threshold (meaning the public-good is provided with smaller ex-ante probability, so less often), and thus lowers the expected social surplus. This theorem also quantifies the tradeoff between gaining privacy (by increasing $\delta$) and losing social surplus (by raising the provision threshold) at a fixed revenue level.

\begin{remark}
In Theorem~\ref{thm:surplus_rev}, we maximize the \textit{level} of the social surplus subject to a revenue constraint at a fixed noise level. In Section~\ref{sec:noise_in_alloc}, we demonstrate that in large economies, the distortion in social surplus closely tracks the corresponding noise sensitivity. Consequently, Theorem~\ref{thm:main} can be interpreted as a program that minimizes the distortions in social surplus while maintaining a revenue constraint at a fixed noise level. Together, these two theorems examine the \textit{level} and \textit{distortions} of social surplus under a revenue constraint across varying levels of privacy-preserving noise.
\end{remark}

The plan in the following sections is to develop the necessary tools for proving the above claims. In the process, we will also provide additional insights and methods for examining the effects of noise on mechanisms' performance, offering a more comprehensive understanding of the topic.

In Section~\ref{sec:implement}, we study the incentive compatibility and individual rationality of the mechanisms in the noisy environment. Then, in Section~\ref{sec:noise_in_alloc} we study the impact of noise in the allocation rule. Specifically, we present a self-contained introduction to the Fourier analysis of Boolean functions, and use it to study the comparative statics of social surplus and revenue with respect to the noise. Having introduced the required tools, we prove the theorems in Section~\ref{sec:thm_proofs}. Finally in Section~\ref{sec:imperfect_know}, we investigate the implications of our methodology in an environment where noise is used to represent agents' imperfect knowledge of their own preferences, rather than as a privacy protection measure.

\section{Implementation under Noise}
In this section, we study the incentive compatibility and individual rationality in the presence of noise. We show how the space of implementable SCFs vary with respect to the noise, and we offer a version of revenue equivalence theorem for any implementable SCF.
\label{sec:implement}
\subsection{Incentive Compatibility}
Suppose individual $i$ reports message $m_i$ to the planner. Denote the received message (that is subject to noise) by $y_i(m_i)$, so that $y_i(m_i) = m_i$ with probability $1-\delta$, and $y_i(m_i)=-m_i$ with probability $\delta$. Also, let us denote the vector of reported types by $m=(m_1,\ldots,m_n)$, and the vector of true types by $x=(x_1,\ldots,x_n)$. A mechanism $(f,t_1,\ldots,t_n)$ is Bayes-Nash incentive compatible (BN-IC), if for every $i\in [n]$ and $x_i\in  \{-1,+1\}$ one has
\begin{equation}
\label{eq:interim_BNIC}
\begin{gathered}
    \left(\frac{b+x_i}{2}\right)\BE\left[f\big(y_i(x_i),y_{-i}(x_{-i})\big) \big| x_i\right] - \BE\left[t_i\big(y_i(x_i),y_{-i}(x_{-i})\big) \big| x_i\right] \geq \\ 
    \left(\frac{b+x_i}{2}\right)\BE\left[f\big(y_i(-x_i),y_{-i}(x_{-i})\big) \big| x_i\right] - \BE\left[t_i\big(y_i(-x_i),y_{-i}(x_{-i})\big) \big| x_i\right]\,,
\end{gathered}
\end{equation}
where the expectations are taken with respect to $x_{-i}$ and their flips, namely $y_{-i}(x_{-i})$, as well as the noise in $y_i(x_i)$. To reduce clutter, we use $y_j$ instead of $y_j(x_j)$, and similarly, $y_{-j}$ instead of $y_{-j}(x_{-j})$. Also, as a shorthand, for every function $g:\{-1,+1\}^n \to \BR$, define $\bar{g}_i(x_i) \coloneq\BE\left[g(x_i,x_{-i}) \big| x_i\right]$.

In the following lemma, we characterize the space of all BN-IC direct mechanisms.
\begin{lemma}[BN-IC]
\label{lem:BNIC}
A mechanism consisting of the allocation rule $f$ and the transfer functions $t=(t_1,\ldots,t_n)$ is Bayes-Nash incentive compatible if and only if for every $i\in [n]$,
\begin{equation}
\label{eq:simplified_BNIC}
    \left(\frac{b+1}{2}\right)\big(\bar{f}_i(+1)-\bar{f}_i(-1)\big) \geq \bar{t}_i(+1)-\bar{t}_i(-1) \geq \left(\frac{b-1}{2}\right)\big(\bar{f}_i(+1)-\bar{f}_i(-1)\big)\,.
\end{equation}
\end{lemma}
A corollary of the previous lemma is that the SCF $f$ is implementable in the Bayes-Nash sense if and only if $\bar{f}_i(+1)-\bar{f}_i(-1)\geq 0$ for all $i\in [n]$. We call this property the \textit{marginal monotonicity} of the allocation rule $f$. The concept of marginal monotonicity simply means on expectation the value of a function increases when the $i$-th input changes from $-1$ to $+1$. 

Another important implication of the previous lemma is that the incentive compatibility of a mechanism does not depend on the noise level $\delta$. In other words, a mechanism is BN-IC in the noisy environment if and only if it is BN-IC in the noise-free setting.


\subsection{Individual Rationality and Expected Revenue}

Suppose that by refusing to participate in the mechanism, any individual can ensure a utility of zero. The mechanism design problem then needs to incorporate interim individual rationality (IIR) constraints:
\begin{equation*}
    \left(\frac{b+x_i}{2}\right)\BE\left[f\big(y_i(x_i),y_{-i}(x_{-i})\big) \big| x_i\right] - \BE\left[t_i\big(y_i(x_i),y_{-i}(x_{-i})\big) \big| x_i\right] \geq 0\,.
\end{equation*}
Employing a similar approach to the one used for the BN-IC constraints, that is taking the expectation with respect to the others' types and noisy flips, one can verify that the above equation reduces to 
\begin{subequations}
\label{eq:simplified_IR}
\begin{align}
    \label{eq:high_type_IR}
    \left(\frac{b+1}{2}\right)\left((1-\delta)\bar{f}_i(+1)+\delta \bar{f}_i(-1)\right) &\geq (1-\delta)\bar{t}_i(+1)+\delta \bar{t}_i(-1)\,,\\
    \label{eq:low_type_IR}
    \left(\frac{b-1}{2}\right)\left(\delta\bar{f}_i(+1)+(1-\delta) \bar{f}_i(-1)\right)&\geq \delta \bar{t}_i(+1)+(1-\delta)\bar{t}_i(-1)
    \,.
\end{align}
\end{subequations}
The first (respectively, second) equation above expresses the IIR condition for the high (respectively, low) type.

By equations~\eqref{eq:simplified_IR}, one notices that the individual rationality constraints are in fact affected by the noise level $\delta$. Since the mechanism can only rely on the noisy reports as the inputs, namely the $y_i$'s, there is always a chance that the message sent by a low type individual flips, and they will end up paying the higher transfer $\bar{t}_i(+1)$ instead of $\bar{t}_i(-1)$ (in the BN sense). Therefore, they need to be compensated for this unexpected flip in order to participate, and this will induce a drag on the space of implementable mechanisms as the noise level increases.

We say a mechanism $(f,t)$ is Bayes-Nash implementable if it is BN-IC and IIR. The next proposition shows that decreasing the noise level weakly \textit{expands} the space of implementable mechanisms.
\begin{proposition}
\label{prop:noise_effect_on_implementation}
Suppose a mechanism $(f,t)$ is Bayes-Nash implementable at the noise level $\delta$. Then, it will remain Bayes-Nash implementable for all $\delta'< \delta$.
\end{proposition}
\begin{proof}
We can express the IIR conditions in~\eqref{eq:simplified_IR} as 
\begin{equation*}
\begin{aligned}
    \left(\frac{b+1}{2}\right)\bar{f}_i(+1)-\bar{t}_i(+1) &\geq \delta \left[ \left(\frac{b+1}{2}\right) \big(\bar{f}_i(+1)-\bar{f}_i(-1)\big)-\big(\bar{t}_i(+1)-\bar{t}_i(-1)\big)\right]\,,\\
    \left(\frac{b-1}{2}\right)\bar{f}_i(-1)-\bar{t}_i(-1) &\geq \delta \left[\big(\bar{t}_i(+1)-\bar{t}_i(-1)\big)-\left(\frac{b-1}{2}\right)\big(\bar{f}_i(+1)-\bar{f}_i(-1)\big)\right]\,.
\end{aligned}
\end{equation*}
The BN-IC constraints in equation~\eqref{eq:simplified_BNIC} imply that the \textit{rhs} to both of the above equations are non-negative. Therefore, decreasing $\delta$ relaxes the inequalities, and hence the claim follows.
\end{proof}

We say a SCF $f:\{-1,+1\}^n\to \{0,1\}$ is Bayes-Nash implementable if there exist transfer rules $t_i:\{-1,+1\}^n\to \BR$ for $i\in [n]$, that make the mechanism $(f,t)$ Bayes-Nash implementable. We now present a \textit{revenue equivalence} type result for implementable SCFs in the current Boolean environment.
\begin{proposition}[Revenue equivalence]
\label{prop:revenue_equiv}
A social choice function $f:\{-1,+1\}^n\to \{0,1\}$ is Bayes-Nash implementable if and only if it satisfies marginal monotonicity. In addition, the maximum expected revenue that the planner can raise from implementing $f$ is 
\begin{equation}
\label{eq:max_exp_rev}
\mathsf{R}_\delta[f] \coloneq (1-2\delta) \BE\left[f(x) \sum_{i=1}^n x_i\right]+\left(\frac{b-1}{2}\right)\BE[f(x)]\,.
\end{equation}
\end{proposition}
Therefore, by marginal monotonicity of an implementable SCF, the first expectation term above is always non-negative, and hence the expected revenue becomes a linearly \textit{decreasing} function in noise. We explore the response of the expected social surplus to the noise level as we introduce further tools in the next section. Finally, the above revenue equivalence representation implies the following result.
\begin{corollary}[Maximum expected revenue]
\label{cor:max_exp_rev_by_majority_rule}
In the space of all implementable Boolean SCFs, the majority rule asymptotically extracts the maximum expected revenue, where
\begin{equation}
\label{eq:majority_rule}
    f_{\text{maj}}(x) = \mathbf{1}\left\{\sum_{i=1}^n x_i \geq 0\right\}\,.
\end{equation}
To see this, note that $\mathsf{R}_\delta[f]$ is linear in $f$, thus the following linear threshold function maximizes the expected revenue:
\begin{equation*}
    \hat{f}_n(x)=\mathbf{1}\left\{\sum_{i=1}^n x_i\geq \frac{1-b}{2(1-2\delta)}\right\}\,.
\end{equation*}
Let us denote the above threshold by $\tau \coloneq\tau(b,\delta)$. The expected revenue associated with this SCF is
\begin{equation*}
    \mathsf{R}_\delta[\hat{f}_n] = (1-2\delta)\BE\left[\sum_{i=1}^n x_i \, \cdot \,\mathbf{1}\left\{\sum_{i=1}^n x_i\geq \tau\right\}\right] + \left(\frac{b-1}{2}\right) \BP\left(\sum_{i=1}^n x_i\geq \tau\right) = \frac{1-2\delta}{\sqrt{2\pi}}\, \sqrt{n}\big(1+o(1)\big)\,,
\end{equation*}
where the last equality follows from the application of the central limit theorem as $n\to \infty$ over the i.i.d.\, random variables $\{x_i: i \in [n]\}$. A similar approach shows that the expected revenue associated with $f_\text{maj}$ is equal to $\frac{1-2\delta}{\sqrt{2\pi}}\, \sqrt{n}\big(1+o(1)\big)$, thus it asymptotically raises the maximum expected revenue. The above analysis is the reason we normalized the target revenue $R$ (in Section~\ref{sec:results}) by $(1-2\delta)\sqrt{n}$.
\end{corollary}

\section{Noise in the Allocation Rule} 
\label{sec:noise_in_alloc}
In Section~\ref{subs:quant}, we introduced the noise sensitivity of a SCF and outlined three reasons for why this quantity is of interest. Here, we elaborate on the second one, which focuses on the distortions in the social surplus caused by the added noise. Specifically, we ask how does the level of noise affect the resulting social surplus of the economy? Denote by $S(x,f)$ the social surplus (namely the individuals' utility plus the revenue raised by the planner) when the true vector of types is $x$ and the implemented outcome is $f\in \{0,1\}$. Formally, it is equal to
\begin{equation}
\label{eq:realized_social_surplus}
    S(x,f) = \sum_{i=1}^n\left(\frac{b+x_i}{2}\right)f\,.
\end{equation}
We define the surplus distortion (denoted by $\mathsf{SD}_\delta[f]$) as the $L^2$ distance between what could have been achieved (i.e., $S(x,f(x))$) and what was ultimately realized (i.e., $S(x,f(y))$) as a result of noisy reports:
\begin{equation*}
    \mathsf{SD}_\delta[f]\coloneq\BE\left[\Big(S(x,f(y))- S(x,f(x))\Big)^2\right]\,.
\end{equation*}
In the next proposition, we show asymptotically as $n\to \infty$, the surplus distortion closely follows the noise sensitivity, and thus providing additional support for minimizing the noise sensitivity in program~\eqref{eq:optimization_prob}. In particular, according to the following proposition, one can assert that in large economies, selecting a SCF that minimizes the surplus distortion is equivalent to minimizing the noise sensitivity.
\begin{proposition}
\label{prop:surplus_dist}
$\lim\limits_{n \to \infty} \left| \frac{1}{n^2} \mathsf{SD}_\delta[f] - \frac{b^2}{4}\, \mathsf{NS}_\delta[f] \right| = 0$, uniformly over all $f:\{-1,+1\}^n\to \{0,1\}$.
\end{proposition}

Motivated by the need to study the comparative statics with respect to the noise level $\delta$, and further studying the notion of noise sensitivity, in the next part, we briefly present a self-contained introduction to the Fourier analysis of Boolean functions. A tool that can be applied extensively to many questions in the Boolean environments \citep[e.g., see its application in social choice][]{kalai2002fourier}.\footnote{The interested reader is encouraged to refer to the book by \cite{o2014analysis} and read further topics in this area.}

\subsection{Fourier Analysis of Boolean Functions}
\label{subs:Fourier_Analysis}
Let the $n$-dimensional Boolean hypercube $\{-1,+1\}^n$ be equipped with the uniform probability measure. The space of $\BR$-valued and square integrable functions on this hypercube, denoted by $H\coloneq L^2\left(\{-1,+1\}^n\right)$, is in fact a separable Hilbert space with the inner product operator:
\begin{equation*}
    \langle f,g\rangle = \BE\left[f(x)g(x)\right] = \frac{1}{2^n}\sum_{x \in \{-1,+1\}^n} f(x)g(x), \quad \quad \forall f,g \in H\,.
\end{equation*}
For every subset $S \subseteq [n]$, define $\chi_S(x) \coloneq \prod_{i \in S}x_i$. It can be readily checked that the collection of functions $\left\{\chi_S(\cdot): S\subseteq [n]\right\}$ constitutes an \textit{orthonormal} basis for $H$. In particular, for $S=\emptyset$, one has $\chi_\emptyset(\cdot)\equiv 1$. Every function $f\in H$ thus has a \textit{unique} Fourier expansion in terms of these basis elements, namely
\begin{equation}
\label{eq:Fourier_expansion}
    f(x) = \sum_{S \subseteq [n]} \hat{f}(S) \chi_S(x)\,,
\end{equation}
in that $\hat{f}(S)$ is called a Fourier coefficient of $f$, and is the projection $f$ onto $\chi_S$, that is 
\begin{equation*}
    \hat{f}(S) = \langle f, \chi_S \rangle  = \BE\left[f(x) \chi_S(x)\right]\,.
\end{equation*}
In particular, $\hat{f}(\emptyset)$ is equal to the mean value of $f$ (i.e., $\BE[f]$), and $\hat{f}(\{i\}) = \BE\left[f(x)x_i\right] = \big(\bar{f}_i(+1)-\bar{f}_i(-1)\big)/2$ is called a \textit{degree}-$1$ Fourier coefficient. 
\begin{example}
Let $f(x) = \max\{x_1,x_2\}$, then one can write $f(x)$ as 
\begin{equation*}
    f(x) = \frac{1}{2}+\frac{1}{2}x_1+\frac{1}{2}x_2-\frac{1}{2}x_1x_2\,,
\end{equation*}
therefore, $\hat{f}(\emptyset)=\hat{f}(\{1\})=\hat{f}(\{2\})=1/2$, $\hat{f}(\{1,2\})=-1/2$ and all other Fourier coefficients are zero.
\end{example}
Next, we introduce the concept of \textit{noise stability} that proves very useful in the analysis of noise sensitivity.
\begin{definition}[Noise stability]
Let $f:\{-1,+1\}^n \to \BR$ belong to $H$. Suppose $y$ is the $\delta$-noisy version of the vector $x \sim \text{Unif}\left(\{-1,+1\}^n\right)$. That is, each $y_i$ is independently distributed from other $y_j$'s and $\BP(y_i\neq x_i)=\delta$. Then, the noise stability of the function $f$ is defined as
\begin{equation}
\label{eq:noise_stab_def}
    \mathsf{Stab}_\delta[f] \coloneq \BE\left[f(x)f(y)\right]\,.
\end{equation}
\end{definition}
From the Fourier expansion in equation~\eqref{eq:Fourier_expansion} one has
\begin{equation}
\label{eq:noise_conditional_exp}
\begin{split}
    \BE\left[f(y) | x\right] &= \sum_{S\subseteq [n]} \hat{f}(S) \prod_{i\in S} \BE\left[y_i | x_i\right]\\
    &= \sum_{S\subseteq [n]} \hat{f}(S) \prod_{i\in S} (1-2\delta)x_i = \sum_{S\subseteq [n]} \hat{f}(S) (1-2\delta)^{|S|} \chi_S(x)\,,
\end{split}    
\end{equation}
where $|S|$ refers to the cardinality of the set $S$. Therefore, an equivalent representation for the noise stability (in terms of the Fourier coefficients) would be
\begin{equation*}
    \mathsf{Stab}_\delta[f] = \sum_{S \subseteq [n]} (1-2\delta)^{|S|} \hat{f}(S)^2\,.
\end{equation*}
This representation suggests that higher-degree Fourier weights have a relatively smaller impact on the stability of any SCF due to their effect being subject to a geometric discounting.

Using the concepts introduced above, in the next part we explore the comparative statics of the social surplus and the revenue with respect to the noise level $\delta$.

\subsection{Impact of Noise on Revenue and Surplus}
We saw in Proposition~\ref{prop:revenue_equiv} that increasing the noise level $\delta$ decreases the expected revenue in every implementable SCF. Now we see how the ideas from spectral analysis offered in the previous section may be directly applied to study the comparative statics of expected social surplus with respect to the noise.

Equation~\eqref{eq:realized_social_surplus} expresses the \textit{realized} social surplus, when the individuals' true type is $x$, and the implemented outcome is $f\in \{0,1\}$. Therefore, in the noisy setting where $f(y)$ is directed instead of $f(x)$, the expected social surplus is equal to 
\begin{equation}
\label{eq:exp_social_surplus}
    \mathsf{S}_\delta[f] \coloneq \BE\left[S(x,f(y))\right] = \sum_{i=1}^n \BE\left[\left(\frac{b+x_i}{2}\right)f(y)\right]\,.
\end{equation}
In the next proposition, we offer the comparative statics of $(\mathsf{R}_\delta,\mathsf{S}_\delta)$ with respect to the noise level $\delta$. Before that, we highlight an important connection between implementability and Fourier coefficients.
\begin{remark}
A SCF $f$ is implementable if and only if all of its degree-$1$ Fourier coefficients are non-negative. This is the case because (Bayes-Nash) implementability is equivalent to marginal monotonicity, and that in turn means $\bar{f}_i(+1)-\bar{f}_i(-1) \geq 0$ for all $i\in [n]$. The former difference is simply equal to $2\hat{f}(\{i\})$, and thus the claim follows.
\end{remark}
\begin{proposition}[Comparative statics]
\label{prop:comparative_statics_noise}
For every implementable SCF $f$, as the noise level $\delta \in (0,1/2)$ increases, the expected revenue $\mathsf{R}_\delta[f]$ and the expected social surplus $\mathsf{S}_\delta[f]$ decrease linearly in $\delta$. 
\end{proposition}
\begin{proof}
It was previously shown in the revenue equivalence expression~\eqref{eq:max_exp_rev} that $\mathsf{R}_\delta$ is a linearly decreasing function of $\delta$. Next, using the expression~\eqref{eq:exp_social_surplus} and the expansion for the conditional expectation in~\eqref{eq:noise_conditional_exp}, one obtains the following representation for $\mathsf{S}_\delta$:
\begin{equation*}
\begin{gathered}
    \mathsf{S}_\delta[f]= \sum_{i=1}^n \BE\left[\left(\frac{b+x_i}{2}\right)f(y)\right] = \sum_{i=1}^n \BE\left[\left(\frac{b+x_i}{2}\right)\sum_{S \subseteq [n]} \hat{f}(S) (1-2\delta)^{|S|} \chi_S(x)\right]\\
    =\frac{b\,n}{2}\,\hat{f}(\emptyset)+\frac{1-2\delta}{2}\,\sum_{i=1}^n \hat{f}(\{i\})\,.
\end{gathered}
\end{equation*}
The last equality holds because $\BE\left[\chi_S(x)\right]=0$ for all $S\neq \emptyset$, and $\BE\left[x_i \chi_S(x)\right]=1$ if $S=\{i\}$ and otherwise is equal to zero. Since $f$ is implementable, then all of its degree-$1$ Fourier coefficients are non-negative, and hence $\mathsf{S}_\delta$ becomes a linearly decreasing function in $\delta$.
\end{proof}
The intuition behind this result is rather simple. As it relates to the expected revenue, a low type agent must be compensated enough to participate, because there is always a chance that their message flips and they end up paying the high type transfer, even though they derive no utility from the public good. The higher the noise level, the more a low type agent ought to be compensated. On the other hand, a positive transfer from the planner to a low type agent seems alluring to a high type individual. Therefore, to deter them from misreporting their type, the planner has to reduce the transfer \textit{paid} by a high type agent. Both of these two effects create a negative pressure on the expected revenue as the noise level increases. For the expected social surplus, observe the complementarity between the outcome $f$ and the agent's type in the utility function (equation~\eqref{eq:utility_form}). Introducing the noise breaks the optimal assortative allocation with some positive probability and thus lowers the expected social surplus.

Proposition~\ref{prop:comparative_statics_noise} also underscores the cost of protecting privacy. While adding noise to individuals' messages protects the full revelation of their private types,  it comes at the cost of decreasing the expected revenue and social surplus associated with each implementable SCF.

In terms of differential privacy, Proposition~\ref{prop:comparative_statics_noise} describes how privacy guarantees translate into efficiency and revenue losses. If we desire an $\ve$-differentially private mechanism, then the relation $\delta=(1+\ee^{\ve})^{-1}$ (followed from~\eqref{eq:epsilon_diff}), and the linear dependence of revenue and surplus on $\delta$, quantify the economic consequences of a given privacy guarantee.

To develop some intuition, in the next section, we provide insights about the revenue and the noise sensitivity of the majority rule. Using them as a stepping stone, we provide the solution to the optimization problem of~\eqref{eq:optimization_prob} and~\eqref{eq:surplus_rev_optimization} in Section~\ref{sec:thm_proofs}.

\subsection{Majority Rule}
\label{subs:majority_rule}
In Corollary~\ref{cor:max_exp_rev_by_majority_rule}, we showed that asymptotically as $n\to \infty$, the majority rule extracts the maximum expected revenue. In the following proposition, we provide representations for its revenue and noise sensitivity.
\begin{proposition}[Majority rule]
\label{prop:maj_rule}
The expected revenue and the noise sensitivity of the majority rule are as follows:
\begin{equation*}
    \begin{split}
        \mathsf{R}_\delta[f_{\text{maj}}] &= \frac{(1-2\delta)}{\sqrt{2\pi}}\sqrt{n}\big(1+o(1)\big)\,,\\
        \mathsf{NS}_\delta[f_\text{maj}] &= \frac{\arccos(1-2\delta)}{\pi} \big(1+o(1)\big)\,.
    \end{split}
\end{equation*}
\end{proposition}

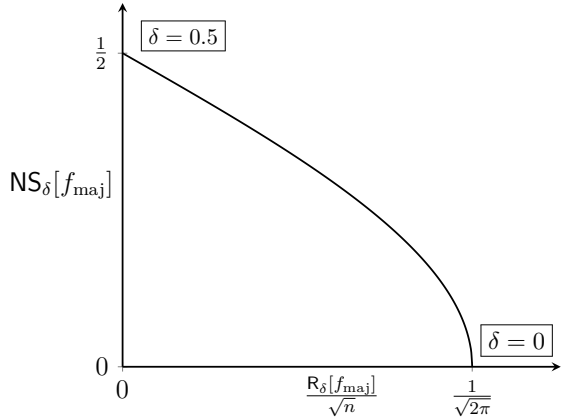
\begin{figure}[htp]
\begin{center}
\begin{tikzpicture}[scale=0.85]
    \begin{axis}[
    axis lines=left,
    smooth,
    axis line style={thick},
    xtick={0,0.398942},
    xticklabels={$0$,$\frac{1}{\sqrt{2\pi}}$},
    xmin=0,
    xmax=0.5,
    ytick={0,0.5},
    yticklabels={$0$,$\frac{1}{2}$},
    ytick pos=left,
    x label style={at={(axis description cs:0.5,0)},anchor=north},
	y label style={at={(axis description cs:0,0.5)},anchor=east,rotate=-90},
    xlabel=$\frac{\mathsf{R}_\delta[f_{\text{maj}}]}{\sqrt{n}}$,
    ylabel=${\mathsf{NS}_\delta [f_{\text{maj}}]}$,
    ymin=0,
    ymax=0.58,
    ]
    \addplot[thick,samples=400,domain= 0:1/sqrt(2*pi),trig format=rad]
    {acos((sqrt(2*pi)*x))/pi} [xshift=+20pt,yshift=8pt]
	node [pos=0.98] {\small{$\boxed{\delta=0}$}}
	[xshift=3pt,yshift=2pt]
	node [pos=0.02] {\footnotesize{$\boxed{\delta=0.5}$}}
	;
    \end{axis}
\end{tikzpicture}
\caption{Revenue and Noise Sensitivity of the Majority Rule}
\label{fig:maj_rule}
\end{center}
\end{figure}

The curve in Figure~\ref{fig:maj_rule} traces the asymptotic values for the \textit{normalized} expected revenue (on the $x$-axis) and the noise sensitivity (on the $y$-axis) of the majority rule as $n\to \infty$, while the noise parameter $\delta$ varies from $0$ to $0.5$. As previously mentioned, higher levels of noise are associated with better privacy protection, higher noise sensitivity, and lower expected revenue for every implementable SCF (and here in particular for $f_{\text{maj}}$). 

A small increase in $\delta$ relative to the noise-free environment changes the expected revenue by a little, but significantly raises the noise sensitivity. This is owed to the fact that expected revenue changes linearly in $\delta$, but the noise sensitivity of the majority rule has ``\textit{infinite}'' derivative at $\delta=0$.

Recall that using the language of differential privacy, $\delta$ is connected to the privacy guarantee of $\ve$. So our results provide a quantitative relation between the promised level of privacy, the resulting noise sensitivity, and revenue loss for the majority function.

A natural question to ask is: given a fixed level of $\delta$, how much noise robustness can be gained if one is willing to sacrifice some revenue relative to the majority function? This question is the basis of the program in~\eqref{eq:optimization_prob}, which we investigate in the next section.

\section{Proof of the Main Results}
\label{sec:thm_proofs}
In this section, we provide the proofs of the main results in Section~\ref{sec:results}.
\subsection{Proof of Theorem~\ref{thm:main}}
\label{subs:main_thm_proof}
In this section, we find the asymptotically optimal solution to the optimization problem of~\eqref{eq:optimization_prob}. Specifically, we ask whether one can find a curve which consistently stays below the one in Figure~\ref{fig:maj_rule}. That is, for a certain level of expected revenue, is there any implementable Boolean function that achieves a smaller noise sensitivity than the majority rule? We answer this question affirmatively and prove that there are two LTFs, whose thresholds are symmetric around 50\%, which are asymptotically optimal for optimization~\eqref{eq:optimization_prob}. The one with the smaller provision threshold has the additional advantage of maximizing the expected social surplus $\mathsf{S}_\delta[\cdot]$ given a target revenue level (this is the content of Theorem~\ref{thm:surplus_rev}).

Our proof follows three main steps: (\rn{1}) simplifying the objective and the constraint set; (\rn{2}) relaxing the constraint set; and (\rn{3}) identifying the asymptotically optimal solutions in the relaxed region and demonstrating that they also belong to the original constraint set.

\subsubsection*{Step (\rn{1})} Observe that since the allocation rules in~\eqref{eq:optimization_prob} are $\{0,1\}$-valued, then
\begin{equation*}
    \mathsf{NS}_\delta[f]=\BP\big(f(x)\neq f(y)\big) = \BE\left[\big(f(x)-f(y)\big)^2\right] = 2\BE\left[f(x)^2\right]-2\BE\left[f(x)f(y)\right]\,.
\end{equation*}
Therefore, we can express the noise sensitivity in terms of the noise stability defined in~\eqref{eq:noise_stab_def}, namely $\mathsf{NS}_\delta[f] = 2\left(\BE[f]-\mathsf{Stab}_\delta[f]\right)$.\footnote{Here, we used the fact that $\BE[f^2]=\BE[f]$ because $f$ is $\{0,1\}$-valued.} Also, note that since the range of all allocation rules is the binary set $\{0,1\}$, the program in~\eqref{eq:optimization_prob} always has a solution.

\begin{remark}
\label{rem:anonymity}
Since the optimization problem~\eqref{eq:optimization_prob} involves no ex ante heterogeneity across the input coordinates $\{x_i:i\in [n]\}$, there is always a solution that respects the \textit{anonymity} of the type vector $x$. Formally, the optimal solution only depends on the number of $+1$'s (or equivalently $-1$'s) in the input vector. Therefore, without any loss, we can restrict the constraint set in this optimization problem to all functions that also satisfy the anonymity condition. Henceforth, with some abuse of notation, we refer to $f(x)$ by $f(\nu_n(x))$, $f(\nu)$, or sometimes $f$.
\end{remark}

Following the remarks in Corollary~\ref{cor:max_exp_rev_by_majority_rule} and Notation~\ref{not:normalization}, we normalize the target revenue, $r\coloneq R/(1-2\delta)\sqrt{n}$. We further assume $r< 1/\sqrt{2\pi}$, as otherwise when $n\to \infty$, there is no SCF (other than the majority rule) that extracts such a high expected revenue.

From revenue equivalence in Proposition~\ref{prop:revenue_equiv}, we know one can always find a set of transfers, that extract the maximum expected revenue from an implementable SCF $f$. Hence, thanks to the anonymity condition the expression~\eqref{eq:max_exp_rev} simplifies to:
\begin{equation*}
    \mathsf{R}_\delta[f] = (1-2\delta)\BE\left[f(\nu_n)\, \nu_n\right]+\left(\frac{b-1}{2}\right)\BE[f(\nu_n)]\,.
\end{equation*}
Finally, recall that a SCF is implementable if and only if it is marginally monotone. Putting the previous derivations together, we can now express an \textit{equivalent} optimization problem to the one in~\eqref{eq:optimization_prob}:
\begin{equation}
    \label{eq:simplified_optimization}
    \begin{gathered}
         \min_f \, \,\, 2\left(\BE[f]-\mathsf{Stab}_\delta[f]\right) \\
         \text{subject to: }\, \frac{1}{\sqrt{n}}\,\BE\left[f(\nu_n)\nu_n\right] + \frac{b-1}{2(1-2\delta)\sqrt{n}}\, \BE\left[f(\nu_n)\right] \geq r,\\
         \text{and } f \text{ being marginally monotone}\,.
    \end{gathered}
\end{equation}
We denote the optimal value of the above minimization problem by $\mathcal{V}_n(r)$, that is equal to the minimum noise sensitivity of the implementable SCFs that raise the normalized expected revenue of $r$.
\begin{remark}
Even if one is willing to convexify the constraint set in~\eqref{eq:simplified_optimization}, by allowing $f$ to take values in the range $[0,1]$, the objective function is still not concave in $f$, and hence the extreme point theory (commonly used in mechanism design literature) cannot be applied.
\end{remark}

\subsubsection*{Step (\rn{2})}
In this part of the proof, we relax the constraint set in~\eqref{eq:simplified_optimization}. Toward this, we \textit{index} the above program with the bias (or the mean) of the SCFs. Specifically, we find the minimum bias of the SCFs that satisfy the above revenue constraint:
\begin{equation}
\label{eq:bias_indexing}
    \alpha_n(r)\coloneq\inf\left\{\BE[f]:  \frac{1}{\sqrt{n}}\BE\left[f(\nu_n)\nu_n\right] + \frac{b-1}{2(1-2\delta)\sqrt{n}}\, \BE\left[f(\nu_n)\right] \geq r, \, \, f\in H_{[0,1]}\right\}\,,
\end{equation}
where $H_{[0,1]}$ is the closed subset of $L^2$ functions from $\{-1,+1\}^n$ to $[0,1]$. Since this is a compact subset, and the revenue constraint induces a closed region, the above infimum is achieved.
\begin{lemma}[Bias indexing]
\label{lem:bias_indexing}
The optimal solution to the minimization problem of~\eqref{eq:bias_indexing} is obtained by $\bar{\ell}_n(\cdot; r)$ in~\eqref{eq:upper_LTF}, and the minimum value $\alpha_n(r)$ satisfies
\begin{equation}
     \label{eq:bias_indexing_optimal_value}
    \lim_{n \to \infty} \alpha_n(r) = \Phi\left(-\varphi^{-1}(r)\right)\,.
\end{equation}
\end{lemma}
This result tells us among all SCFs that raise a target revenue the \textit{linear threshold functions} have the smallest mean. In addition, the associated threshold depends on the normalized revenue $r$. Higher levels of normalized revenue corresponds to smaller thresholds, thus getting closer to the majority rule.
\begin{remark}
Taking $r<1/\sqrt{2\pi}$, equation~\eqref{eq:bias_indexing_optimal_value} implies that $\lim_{n\to \infty} \alpha_n(r)<1/2$, so for all $n$ greater than a certain level, one has $\alpha_n(r)<1/2$. Therefore, we can define the \textit{mirrored} optimization problem to the one in~\eqref{eq:bias_indexing} as follows:
\begin{equation}
\label{eq:mirrored_bias_indexing}
    \sup\left\{\frac{1}{\sqrt{n}}\,\BE\left[f \nu_n\right]+\frac{b-1}{2(1-2\delta)\sqrt{n}}\, \BE\left[f\right]: \BE[f]\geq 1-\alpha,\,\, f\in H_{[0,1]}\right\}\,.
\end{equation}
Since the distribution of $\nu_n$ is symmetric around $0$, one can see that there exists an $o(1)$ sequence such that, replacing $\alpha$ with $\alpha_n(r)+o(1)$ in the above constraint leads to a supremum of $r$. That is 
\begin{equation}
\label{eq:mirrored_b}
    \BE[f]\geq 1-\big(\alpha_n(r)+o(1)\big) \text{ implies } \frac{1}{\sqrt{n}}\,\BE\left[f \nu_n\right]+\frac{b-1}{2(1-2\delta)\sqrt{n}}\, \BE\left[f\right]\leq r\,.
\end{equation}
\end{remark}
And specifically, using the same techniques as in Lemma~\ref{lem:bias_indexing}, one can show the existence of a LTF with the following description (as previously expressed in equation~\eqref{eq:lower_LTF}),
\begin{equation*}
    \underline{\ell}_n(x;r) \coloneq \mathbf{1}\left\{\frac{\nu_n(x)}{\sqrt{n}} \geq -\varphi^{-1}(r)+o(1)\right\}\,,
\end{equation*}
that exactly achieves the normalized revenue $r$, and its mean, i.e., $\BE\left[\underline{\ell}_n(x;r)\right]$ equals $1-\big(\alpha_n(r)+o(1)\big)$.

Our next step is to use the idea of bias indexing to relax the constraint set in~\eqref{eq:simplified_optimization}. Observe that the definition of $\alpha_n(\cdot)$ in~\eqref{eq:bias_indexing} and the condition~\eqref{eq:mirrored_b} jointly imply the following set inclusion:
\begin{equation*}
\begin{gathered}
    \left\{f\in H_{[0,1]}: \frac{1}{\sqrt{n}}\,\BE\left[f \nu_n\right]+\frac{b-1}{2(1-2\delta)\sqrt{n}}\, \BE\left[f\right] \geq r \text{ and } f \text{ being marginally monotone}\right\}\\
    \subseteq \Big\{f\in H_{[0,1]}: \alpha_n(r) \leq \BE\left[f\right] \leq 1- \big(\alpha_n(r)+o(1)\big)\Big\}\,.
\end{gathered}
\end{equation*}
Consequently, we can relax the constraint set of the original problem and arrive to the following program:
\begin{equation}
\label{eq:relaxed_optimization}
\begin{gathered}
    \min_f \, \,  2\left(\BE[f]-\mathsf{Stab}_\delta[f]\right)\\
    \text{subject to: } \, \alpha_n(r) \leq \BE\left[f\right] \leq 1- \big(\alpha_n(r)+o(1)\big) \text{ and } f \in H_{[0,1]}\,.
\end{gathered}
\end{equation}
We denote the value to this minimization problem by $\mathcal{V}_n^{\text{rel}}(r)$, and importantly we have $\mathcal{V}_n^{\text{rel}}(r) \leq \mathcal{V}_n(r)$.

\subsubsection*{Step (\rn{3})}
In this part, we characterize the asymptotically optimal solutions for the relaxed program, and further show they satisfy the original constraint in~\eqref{eq:simplified_optimization}, thereby proving their asymptotic optimality for the main program.

Essentially, we will show that the LTFs $\bar{\ell}_n(\cdot;r)$ (in equation~\eqref{eq:upper_LTF}) and $\underline{\ell}_n(\cdot;r)$ (in equation~\eqref{eq:lower_LTF}) are approximately optimal for the relaxed program. The former achieves the bias lower bound in~\eqref{eq:relaxed_optimization}, and the latter achieves the bias upper bound. Additionally, since both functions satisfy the constraints of the original optimization problem in~\eqref{eq:simplified_optimization}---namely raising precisely $r$ and being marginally monotone---they will remain asymptotically optimal for the original program. The approximation error due to choosing them as suboptimal solutions for~\eqref{eq:simplified_optimization} converges to zero as $n\to \infty$. 

To justify the previous claims, we borrow from a seminal result in the analysis of Boolean functions, that goes under the name of ``majority is the stablest'', and its proof mainly relies on the Gaussian isoperimetric inequality (first proved by~\cite{borell1985geometric}). In the following lemma we present a version of this result that suits our need, and we provide a rough sketch of its proof in the appendix.\footnote{The original proof is rather long, and has several steps. The curious reader should consult~\cite{Mossel2010noise} or chapter 11.7 of~\cite{o2014analysis} for the complete proof.} Before that we need to define a notation for the two dimensional CDF of correlated Gaussians.
\begin{definition}
\label{def:Gaussian_funcs}
Let $(Z_1,Z_2)$ be two standard Gaussian random variables, that are $\rho$-correlated, namely $\BE\left[Z_1 Z_2\right]=\rho$. We define $\Phi_\rho: \BR^2 \to [0,1]$ as $\Phi_\rho(t_1,t_2) \coloneq \BP_\rho\left(Z_1\leq t_1, Z_2 \leq t_2\right)$. In particular, when $t_1=t_2=t$, with some abuse of notation we use $\Phi_\rho(t)\equiv \Phi_\rho(t,t)$.
\end{definition}
\begin{lemma}[``Majority is the stablest'']
\label{lem:maj_stablest}
Let $f:\{-1,+1\}^n\to [0,1]$ be an anonymous function, and $\delta \in (0,1/2)$, then
\begin{equation*}
    \mathsf{Stab}_{\delta}[f] \leq \Phi_{1-2\delta}\big(\Phi^{-1}(\BE[f])\big)+o(1)\,,
\end{equation*}
where the $o(1)$ approximation term is uniform across all anonymous functions.
\end{lemma}
By using the above inequality, we demonstrate that the two LTFs mentioned earlier are asymptotically optimal for the original program. This establishes the bounds in~\eqref{eq:main}, thereby concluding the proof of Theorem~\ref{thm:main}.

By the previous lemma, the objective function in the relaxed program of~\eqref{eq:relaxed_optimization} is lower bounded by
\begin{equation*}
    \BE[f]-\mathsf{Stab}_\delta[f] \geq \BE[f]-\Phi_{1-2\delta}\big(\Phi^{-1}(\BE[f])\big)+o(1)\,,
\end{equation*}
where the $o(1)$ term is uniform across all anonymous SCFs. The expression on the \textit{rhs} above---up to the exclusion of the $o(1)$ term---is symmetric around $\BE[f]=1/2$. In particular, it is increasing (respectively, decreasing) on the region where $\BE[f] \leq 1/2$ (respectively, $\BE[f]\geq 1/2$).\footnote{Let us define $\Psi(x)\coloneq x -\Phi_{\rho}\big(\Phi^{-1}(x)\big)$ for $x\in [0,1]$. In Appendix~\ref{app:map}, we show that for any $\rho\in [-1, 1]$, the mapping $\Psi$ is increasing on $[0,1/2]$ and decreasing on $[1/2, 1]$.} Therefore, for any anonymous $f$ that belongs to the constraint set of the relaxed problem in~\eqref{eq:relaxed_optimization}, one has
\begin{equation*}
    \BE[f]-\Phi_{1-2\delta}\big(\Phi^{-1}(\BE[f])\big) \geq \alpha_n(r)-\Phi_{1-2\delta}\big(\Phi^{-1}(\alpha_n(r)\big)+o(1)\,,
\end{equation*}
where the inequality binds for $f\in \{\underline{\ell}_n(\cdot;r),\bar{\ell}_n(\cdot;r)\}$, because as constructed in step (\rn{2}) above, we have $\BE\left[\bar{\ell}_n(\nu_n;r)\right]=\alpha_n(r)$ and $\BE\left[\underline{\ell}_n(\nu_n;r)\right]=1-\big(\alpha_n(r)+o(1)\big)$. This in turn implies that
\begin{equation*}
    \mathsf{NS}_\delta[\ell_n] \leq \mathcal{V}_n^{\text{rel}}(r)+o(1),\,\, \text{ for } \ell_n \in \{\underline{\ell}_n(\cdot;r),\bar{\ell}_n(\cdot;r)\},\,
\end{equation*}
and hence the second inequality in equation~\eqref{eq:main} follows because $\mathcal{V}_n^{\text{rel}}(r) \leq \mathcal{V}_n(r)$. The first inequality readily holds because $\{\underline{\ell}_n(\cdot;r),\bar{\ell}_n(\cdot;r)\}$ also belong to the constraint set of the original problem in~\eqref{eq:simplified_optimization}, as they both raise the normalized expected revenue of $r$ and are monotone functions. This completes the justification of~\eqref{eq:main}, and hence the proof of Theorem~\ref{thm:main}.\qed


\subsection{Proof of Theorem~\ref{thm:surplus_rev}}
By borrowing the expression found for the expected social surplus in the proof of Proposition~\ref{prop:comparative_statics_noise}, and following the approach in the previous section to simplify the revenue constraint, we can recast the optimization problem of~\eqref{eq:surplus_rev_optimization} as:
\begin{equation}
\label{eq:simplified_surplus_rev_optimization}
\begin{gathered}
     \max \left\{\frac{b}{2}\,\BE[f(\nu_n)]+\frac{1-2\delta}{2n}\,\BE\left[f(\nu_n)\nu_n\right]\right\}\\
     \text{subject to: }\, \frac{1}{\sqrt{n}}\BE\left[f(\nu_n)\nu_n\right] + \frac{b-1}{2(1-2\delta)\sqrt{n}}\, \BE\left[f(\nu_n)\right] \geq r\,,\\
     \text{ and } f \text{ being implementable}\,.
\end{gathered}
\end{equation}
This problem falls under the class of linear programs, in that one needs to assign the optimal value to $f(\nu)$ for every $\nu \in \left\{-n,-n+2,\ldots,n\right\}$. The corresponding Lagrangian for the relaxed problem, where we skip the implementability condition, is
\begin{equation*}
\begin{gathered}
    \mathcal{L} = \frac{b}{2}\BE[f(\nu_n)]+\frac{1-2\delta}{2n}\,\BE\left[f(\nu_n)\nu_n\right] \\
    +\lambda \left(\frac{1}{\sqrt{n}}\,\BE\left[f(\nu_n)\nu_n\right] + \frac{b-1}{2(1-2\delta)\sqrt{n}}\, \BE\left[f(\nu_n)\right]-r\right)\,.
\end{gathered}
\end{equation*}
Since the Lagrange multiplier $\lambda$ is non-negative, then $\lambda+ \frac{1-2\delta}{2\sqrt{n}}>0$, and the candidate solution takes the following form
\begin{equation*}
    f(\nu_n) = \mathbf{1}\left\{\frac{\nu_n}{\sqrt{n}} \geq \frac{-\left(b+\frac{\lambda(b-1)}{(1-2\delta)\sqrt{n}}\right)}{2\lambda+\frac{1-2\delta}{\sqrt{n}}}\right\}\,.
\end{equation*}
One can easily check that increasing $\lambda$ in the above expression, raises the provision threshold, thus asymptotically (as $n\to \infty$) decreases the expected social surplus, while increasing the expected revenue. This is so because the second term in $\mathsf{S}_\delta[f]$ is of order $O(1/\sqrt{n})$ and asymptotically vanishes compared to the first term, which in turn is decreasing in the provision threshold. Therefore, we should find the minimum $\lambda$ that satisfies the revenue constraint. For this, let us denote the threshold by
\begin{equation*}
    \xi_n\equiv \xi_n(b,\delta,\lambda)\coloneq\frac{-\left(b+\frac{\lambda(b-1)}{(1-2\delta)\sqrt{n}}\right)}{2\lambda+\frac{1-2\delta}{\sqrt{n}}}\,.
\end{equation*}
Hence, we seek the minimum $\lambda$ satisfying the following inequality:
\begin{equation*}
    \BE\left[\frac{\nu_n}{\sqrt{n}}\, \cdot \, \mathbf{1}\left\{\frac{\nu_n}{\sqrt{n}} \geq \xi_n(b,\delta,\lambda)\right\} \right]+\frac{(b-1)}{2(1-2\delta)\sqrt{n}}\,\BP\left(\frac{\nu_n}{\sqrt{n}} \geq \xi_n(b,\delta,\lambda) \right) \geq r\,.
\end{equation*}
As $n\to \infty$, the normalized sum $\nu_n/\sqrt{n}$ converges in distribution to the standard Gaussian, thus the \textit{lhs} in the above inequality converges. Specifically, the first term is asymptotically equal to $\varphi(\xi_n)+o(1)$, and the second term is also of $o(1)$. Therefore, the $\lambda$ in $\xi_n$ must be chosen so that
\begin{equation*}
    \varphi\left(\xi_n\right) + o(1) = r\,.
\end{equation*}
Since the provision threshold $\xi_n$ is negative, then the above condition implies that the optimal threshold is $-\varphi^{-1}(r)+o(1)$. Specifically, letting this $o(1)$ sequence be equal to the one in the threshold of $\underline{\ell}_n$ raises precisely the normalized revenue of $r$, thereby verifying the optimality of the LTF in~\eqref{eq:lower_LTF}. \qed

\subsection{Additional Remarks}
\begin{remark}
The normalized expected revenue raised by $\{\underline{\ell}_n(\cdot;r),\bar{\ell}_n(\cdot;r)\}$ is equal to $r$, hence the (unnormalized) expected revenue is $(1-2\delta)\sqrt{n}r$. Furthermore, since $\BE\left[\ell_n \right] = \Phi\big(-\varphi^{-1}(r)\big)+o(1)$, for $\ell_n\in \{\underline{\ell}_n(\cdot;\bar{r}),\bar{\ell}_n(\cdot;\bar{r})\}$, then the noise sensitivity takes the following form:
\begin{equation*}
\begin{split}
    \mathsf{NS}_\delta[\ell_n] &=2\Big\{ \BE\left[\ell_n\right]- \Phi_{1-2\delta}\big(\Phi^{-1}(\BE\left[\ell_n\right])\big)\Big\}\\
    &=2\Big\{\Phi\big(-\varphi^{-1}(r)\big)-\Phi_{1-2\delta}\big(-\varphi^{-1}(r)\big)\Big\}+o(1)\,.
\end{split}
\end{equation*}
We used the above expression to plot the asymptotic Pareto frontiers in Figure~\ref{fig:pareto_frontier}.
\end{remark}

\begin{remark}
    In public good mechanisms, one could envision three main objectives: revenue, surplus, and noise robustness (equivalently, privacy). In our two main theorems, we studied the tradeoffs between each of the last two with the revenue. However, one may question the interaction between social surplus and noise robustness. In fact, in the absence of any revenue constraint, there will be no tradeoff between those two, because the SCF that always provides the public good, achieves the maximum social surplus and zero noise sensitivity. This is so because the per-capita surplus (as stated in the objective function of~\eqref{eq:simplified_surplus_rev_optimization}) consists of two terms. The second component is of $O(1/\sqrt{n})$, and thus negligible compared to the first term. Therefore, by always providing the public good, the first term of the objective function achieves its maximum and the noise sensitivity is zero.
\end{remark}

\section{Imperfect Knowledge of Preferences}
\label{sec:imperfect_know}
So far, we have studied a setting in which individuals perfectly know their preferences, and the noisy flips take place when they send their messages to the planner. We associated two interpretations with this setting: (\rn{1}) noise is deliberately added for privacy-preserving concerns; (\rn{2}) miscommunication between individuals and the planner is inevitable and reported types could alter as a result.

In this section, we turn to an interpretation of our model where individuals simply do not know their own preferences and observe a noisy signal instead. The idea that agents have imperfect knowledge of their own preferences has received some attention in the mechanism design literature, including the recent work of~\cite{gleyze2023} and~\cite{thereze2023}. 
Now, individuals' reported preferences may differ from their true underlying types, not necessarily due to strategic considerations, but because they lack perfect knowledge about their type. Formally, let $x_i$ be uniformly distributed on $\{-1,+1\}$, representing the true type of agent $i$, that is hidden from the agent. Instead, they receive a noisy signal $y_i \in \{-1,+1\}$ that is correlated with their type, in the sense that $\BP\left(y_i=x_i\right)=1-\delta$ for $\delta \in (0,1/2)$. This means the probability that the agent's signal (information) matches their true type is higher than the probability that it differs. As before, we assume the pairs $\{(x_i,y_i): i\in [n]\}$ are independently distributed and each has the same distribution explained before. 

A mechanism $(f,t)$ is Bayes-Nash incentive compatible in this setting, when each agent reports their signal (i.e., $y_i$) truthfully, while taking expectations with respect to the others' types. Let $y=(y_1,\ldots,y_n)$ be the vector of signals transmitted by the individuals. Then, in a BN-IC mechanism the planner outputs $f(y)$ and charges agent $i$ by the amount $t_i(y)$ for each $i \in [n]$. The interim incentive constraint for agent $i$ with signal $y_i$ is:
\begin{equation}
\label{eq:BNIC_imp_know}
\begin{gathered}
    \BE\left[\left(\frac{b+x_i}{2}\right)f(y_i,y_{-i}) \big| y_i\right]-\BE\left[t_i(y_i,y_{-i}) \big| y_i\right] \geq \\
    \BE\left[\left(\frac{b+x_i}{2}\right)f(-y_i,y_{-i}) \big| y_i\right]-\BE\left[t_i(-y_i,y_{-i}) \big| y_i\right]\,.
\end{gathered}
\end{equation}
\begin{lemma}
In the present setting, where agents do not have perfect knowledge about their types, a mechanism $(f,t)$ is BN-IC if and only if for every $i\in [n]$,
\begin{equation}
\label{eq:simplified_BNIC_imp_know}
    \left(\frac{b+1}{2}-\delta\right)\big(\bar{f}_i(+1)-\bar{f}_i(-1)\big) \geq \bar{t}_i(+1)-\bar{t}_i(-1) \geq \left(\frac{b-1}{2}+\delta\right)\big(\bar{f}_i(+1)-\bar{f}_i(-1)\big)\,.
\end{equation}
\end{lemma}
We skip the proof of this lemma. It follows directly from equation~\eqref{eq:BNIC_imp_know}, observing that because of the independence, the conditional distribution of $y_{-i}$ given $y_i$ is the same as the unconditional distribution of $x_{-i}$. The first (respectively, second) inequality in~\eqref{eq:simplified_BNIC_imp_know} refers to the interim IC constraint when $y_i=+1$ (respectively, $y_i=-1$). In a sharp contrast with the previous setting, where noise came around in the communication stage, the incentive constraints are now affected by the noise level $\delta$. This is so because in the former case, the noise could flip the individual's message and alter their
expected transfer, but in the current setting when the agent sends their signal $y_i$, the transfer they expect, namely $\bar{t}_i(y_i)$, is not further modified by the noise. Finally, equation~\eqref{eq:simplified_BNIC_imp_know} also confirms that as the noise level $\delta$ increases the space of BN-IC mechanisms shrinks.

Next, we express the interim individual rationality constraint for the agent $i$ who received the signal $y_i$, and has an outside option of zero:
\begin{equation*}
    \BE\left[\left(\frac{b+x_i}{2}\right)f(y_i,y_{-i}) \big| y_i\right]-\BE\left[t_i(y_i,y_{-i}) \big| y_i\right] \geq 0\,.
\end{equation*}
One can simplify this constraint into two inequalities, that respectively indicate the IR conditions for the high (i.e., $y_i=+1$) and low (i.e., $y_i=-1$) signals:
\begin{subequations}
\begin{align}
    \label{eq:high_type_IR_imp_know}
    \left(\frac{b+1}{2}-\delta\right) \bar{f}_i(+1) &\geq \bar{t}_i(+1)\,,\\
    \label{eq:low_type_IR_imp_know}
    \left(\frac{b-1}{2}+\delta\right) \bar{f}_i(-1) &\geq \bar{t}_i(-1)\,.
\end{align}
\end{subequations}
We now state the counterpart of Proposition~\ref{prop:revenue_equiv} in the current setting.
\begin{proposition}
\label{prop:revenue_equiv_imp_know}
In the present setting, where agents do not have perfect knowledge about their types, a SCF $f:\{-1,+1\}^n\to \{0,1\}$ is implementable if and only if it satisfies marginal monotonicity, namely $\bar{f}_i(+1)-\bar{f}_i(-1)\geq 0$ for all $i\in [n]$. In addition, the maximum expected revenue that the planner can collect from implementing $f$ is 
\begin{equation}
\label{eq:max_exp_rev_imp_know}
\widetilde{\mathsf{R}}_\delta[f] \coloneq (1-2\delta) \BE\left[f(x) \sum_{i=1}^n x_i\right]+\left(\frac{b-1}{2}+\delta\right)\BE[f(x)]\,,
\end{equation}
where the expectation is taken with respect to the uniform measure on $\{-1,+1\}^n$.
\end{proposition}

We continue by studying the revenue/surplus tradeoff when agents have imperfect knowledge of their preferences. Observe that, in the new setting the implemented outcome is $f(y)$ while the agents' true vector of types is $x$. Therefore, the expected social surplus follows the same expression of equation~\eqref{eq:exp_social_surplus}. 
Hence, the revenue/surplus tradeoff is pinned down by the following program:
\begin{equation*}
    \begin{gathered}
        \max \mathsf{S}_\delta[f] \\
        \text{subject to: }\, \widetilde{\mathsf{R}}_\delta[f] \geq R, \text{ and } f \text{ being implementable}\,.
    \end{gathered}
\end{equation*}
As before, we normalize the lower bound on the expected revenue by $r= R/(1-2\delta)\sqrt{n}$. Then, using the same apparatus as in the proof of Theorem~\ref{thm:surplus_rev}, one can show that the same LTF, namely $\underline{\ell}_n(\cdot;r)$, solves the above problem. 

Suppose the required revenue $R$ remains fixed, and one looks at the response of the constrained efficient allocation rule in the above problem to the noise. As the agents' information about their preferences deteriorate (corresponding to an increase in $\delta$), the normalized revenue $r$ increases, and correspondingly the provision threshold gets closer to the simple 50\% majority rule from \textit{below}. Conversely, an improvement in the agents' knowledge about their types, decreases the threshold and thus increases the chances of provision. This means in the societies where agents have better knowledge about their preferences for public good, the expected likelihood of provision in the efficient allocation rule is higher.

Next, we study the revenue/noise robustness tradeoff. Specifically, we ask the similar question expressed in the optimization problem of~\eqref{eq:optimization_prob}, in that one seeks the SCF with the minimum noise sensitivity subject to raising a target level of expected revenue, in the present setting where agents have imperfect knowledge of their types:
\begin{equation}
\label{eq:NS_min_imp_know}
\begin{gathered}
     \min \mathsf{NS}_\delta[f]\\
     \text{subject to: }\, \widetilde{\mathsf{R}}_\delta[f] \geq R \text{ and } f \text{ being implementable}\,.
\end{gathered}
\end{equation}
A quick inspection on the expressions for expected revenue in these two settings, namely equations~\eqref{eq:max_exp_rev} and~\eqref{eq:max_exp_rev_imp_know}, implies that
\begin{equation*}
    \frac{1}{(1-2\delta)\sqrt{n}} \left|\widetilde{\mathsf{R}}_\delta[f]- \mathsf{R}_\delta[f]\right| = o\left(\frac{1}{\sqrt{n}}\right)\,.
\end{equation*}
Therefore, one can follow the same steps taken in Section~\ref{subs:main_thm_proof}, and show that the two LTFs with approximate thresholds (up to $o(1)$ variations) at $-\varphi^{-1}(r)$ and $\varphi^{-1}(r)$ are asymptotically optimal for the above problem. Hence the following proposition---which is the analogue of Theorem~\ref{thm:main} in this setting---follows:
\begin{proposition}
In the present setting, where agents do not have perfect knowledge about their types, the following LTFs are asymptotically optimal for the program in~\eqref{eq:NS_min_imp_know}:
\begin{equation*}
    g_n(x;r)\coloneq \mathbf{1}\left\{\frac{\nu_n(x)}{\sqrt{n}} \geq -\varphi^{-1}(r)+o(1)\right\}\,, \text{ and }  
    h_n(x;r)\coloneq \mathbf{1}\left\{\frac{\nu_n(x)}{\sqrt{n}} \geq \varphi^{-1}(r)+o(1)\right\}\,.
\end{equation*}
\end{proposition}
Quite naturally, the noise sensitivity of the optimal SCF increases as the agents' knowledge of their preferences deteriorate. But more importantly, similar to the interpretation we attached to Figure~\ref{fig:pareto_frontier}, the worse are the agents' knowledge about their preferences (equivalently the higher is $\delta$), the \textit{smaller} expected revenue the planner has to give up in order to gain a certain level of noise robustness.
\section{Conclusion}
We have studied the tradeoffs between privacy preservation, the standard economic objectives of efficiency and revenue, and the stability of the public-good decision rule. Privacy preservation compromises the pursuit of other objectives, but in a large economy we are able to characterize the asymptotically optimal decision rules, and uncover the underlying quantitative tradeoffs.

Our model is standard, but stylized, assuming binary types and a yes/no decision on the provision of a public good. This structure is essential to our results, mainly because we rely on the Fourier analysis of Boolean functions, that is tailored for binary structures. There are some developments about the spectral analysis and noise sensitivity in more general domains (e.g., Chapter~8 of~\cite{o2014analysis}). However, they are not as crisp and sharp as the binary setting, and we think much more work required in this area.

It is natural to ask the questions of our paper in other environments. Preservation of privacy is an overarching concern, and one can imagine private goods models, as well as public-good settings that are richer than the ones we have focused on here, in which to analyze the effect of privacy-preserving noise. We can only hope that our paper proves a useful starting point for further work.


\newpage
\appendix
\section{Proofs}
\label{sec:proofs}
\subsection{Proof of Lemma~\ref{lem:BNIC}}
When the true type of agent $i$ is $x_i$, the incentive constraint in equation~\eqref{eq:interim_BNIC} reduces to
\begin{equation*}
\begin{gathered}
    \left(\frac{b+x_i}{2}\right)\BE\left[f\big(y_i(x_i),y_{-i}\big)\big| x_i \right]-\BE\left[t_i\big(y_i(x_i),y_{-i}\big)\big| x_i\right] \geq \\ \left(\frac{b+x_i}{2}\right)\BE\left[f\big(y_i(-x_i),y_{-i}\big)\big| x_i\right]-\BE\left[t_i\big(y_i(-x_i),y_{-i}\big)\big| x_i\right]\,.
\end{gathered}
\end{equation*}
Since the flips are independent across the individuals, the joint distribution of $\big(y_i(x_i),y_{-i}\big)$ is the same as $\big(y_i(x_i),x_{-i}\big)$. Therefore, one can summarize the previous condition as
\begin{equation*}
\begin{gathered}
    \left(\frac{b+x_i}{2}\right)\BE\left[\bar{f}_i\big(y_i(x_i)\big)\big| x_i\right]-\BE\left[\bar{t}_i\big(y_i(x_i)\big)\big| x_i\right] \geq \\ \left(\frac{b+x_i}{2}\right)\BE\left[\bar{f}_i\big(y_i(-x_i)\big)\big| x_i\right]-\BE\left[\bar{t}_i\big(y_i(-x_i)\big)\big| x_i\right]\,,
\end{gathered}
\end{equation*}
in that the expectation operators only refer to the noisy flips. When $x_i=+1$, we expand this expression and cancel the appearing term $1-2\delta$ from both sides, thereby showing the first inequality constraint in equation~\eqref{eq:simplified_BNIC}. Similarly, when $x_i=-1$, the incentive constraint reduces to the second inequality in~\eqref{eq:simplified_BNIC}.\qed

\subsection{Proof of Proposition~\ref{prop:revenue_equiv}}
We divide the proof into two parts: (\rn{1}) showing the equivalence between marginal monotonicity and Bayes-Nash implementability; (\rn{2}) proving the revenue equivalence representation in equation~\eqref{eq:max_exp_rev}.

\noindent \textbf{Part (\rn{1})}: As a rather immediate corollary of incentive constraints in~\eqref{eq:simplified_BNIC}, one can observe that the marginal monotonicity of SCF is necessary for every BN-IC mechanism $(f,t)$. It is further sufficient, because if $\bar{f}_i(+1)-\bar{f}_i(-1)\geq 0$ for all $i\in [n]$, one can always find a set of transfer functions, $t=(t_1,\ldots,t_n)$, such that their induced marginals $\big(\bar{t}_i(-1),\bar{t}_i(+1)\big)$ satisfy the BN-IC condition in equation~\eqref{eq:simplified_BNIC}, and the two IIR conditions in~\eqref{eq:simplified_IR} for each $i\in [n]$. To justify this claim, let $\big(\bar{t}_i(-1),\bar{t}_i(+1)\big) = (\beta_{-1},\beta_{+1})$ be any pair that satisfies the BN-IC condition of equation~\eqref{eq:simplified_BNIC} and the IIR conditions of~\eqref{eq:simplified_IR}, induced by the marginally monotone pair $\big(\bar{f}_i(-1),\bar{f}_i(+1)\big)$. We want to show that there exists a function $t:\{-1,+1\}^n\to \BR$, whose marginals on the $i$-th coordinate (averaging out other coordinates) match $(\beta_{-1},\beta_{+1})$. To find such a function, we restrict the search to the smaller space of ``anonymous'' functions, whose value only depend on the number of $+1$'s in the input vector, namely on 
\begin{equation*}
    m(x) \coloneq \#\left\{i: x_i=+1\right\}\,.
\end{equation*}
Therefore, we denote $t(x)$ by $t(m(x))$. Hence, it is required that
\begin{equation*}
    \begin{aligned}
        \beta_{-1} &= \sum_{m=0}^{n-1} t(m)\, \binom{n-1}{m}\frac{1}{2^{n-1}}\,,\\
        \beta_{+1} &= \sum_{m=0}^{n-1} t(m+1)\, \binom{n-1}{m}\frac{1}{2^{n-1}}\,.
    \end{aligned}
\end{equation*}
Let us denote the anonymous function $t(\cdot)$ by the vector $\bm{t} \equiv \big(t(0),t(1),\ldots,t(n)\big)$, and $\binom{n}{k}$ by $C_{n,k}$. Then, the above linear system is expressed by
\begin{equation*}
    \begin{bmatrix}
    C_{n-1,0} & C_{n-1,1} & \ldots & C_{n-1,n-1} & 0\\
    0 & C_{n-1,0} & C_{n-1,1} & \ldots & C_{n-1,n-1} 
    \end{bmatrix}\bm{t} = 2^{n-1}\begin{bmatrix}
    \beta_{-1}\\\beta_{+1}
    \end{bmatrix}\,.
\end{equation*}
Since, the first and last columns of the coefficient matrix are linearly independent, then there always exists a solution to the above system. Therefore, one can always find an anonymous transfer function $t_i(\cdot)$ that implements the marginally monotone pair $\big(\bar{f}_i(-1),\bar{f}_i(+1)\big)$.
\medskip

\noindent \textbf{Part (\rn{2}):} For any implementable SCF $f$, the planner receives the expected transfer
\begin{equation}
\label{eq:exp_transfer_individual_i}
    \frac{1}{2}\big(\bar{t}_i(+1)+\bar{t}_i(-1)\big)\,,
\end{equation}
from individual $i$. Therefore, one should maximize this expression, subject to the BN-IC and IIR conditions, to achieve the maximum expected transfer obtained from the SCF $f$. To solve this program, we first show the IIR condition for the low type (namely equation~\eqref{eq:low_type_IR}) together with the BN-IC condition for the high type (namely the first inequality in~\eqref{eq:simplified_BNIC}) imply the IIR condition for the high type (i.e., equation~\eqref{eq:high_type_IR}). From the low type IIR condition one obtains
\begin{equation}
\label{eq:t_minus_upper_bound}
    \bar{t}_i(-1) \leq -\frac{\delta}{1-\delta}\,\bar{t}_i(+1)+\left(\frac{b-1}{2}\right)\left(\frac{\delta}{1-\delta}\, \bar{f}_i(+1)+\bar{f}_i(-1)\right)\,,
\end{equation}
and the high type BN-IC condition implies
\begin{equation*}
    \bar{t}_i(+1) \leq \bar{t}_i(-1)+\left(\frac{b+1}{2}\right)\left(\bar{f}_i(+1)-\bar{f}_i(-1)\right)\,.
\end{equation*}
Replacing the former upper bound on $\bar{t}_i(-1)$ in the above inequality and applying some rearrangements imply that
\begin{equation}
\label{eq:t_plus_upper_bound}
    \bar{t}_i(+1) \leq \left(\frac{b+1}{2}-\delta\right)\bar{f}_i(+1)-(1-\delta)\bar{f}_i(-1)\,.
\end{equation}
Next, let us investigate the validity of the high type IIR condition (equation~\eqref{eq:high_type_IR}). We use equations~\eqref{eq:t_minus_upper_bound} and~\eqref{eq:t_plus_upper_bound} to obtain the following upper bound on the expected transfer paid by the high type, namely the \textit{rhs} of equation~\eqref{eq:high_type_IR}:
\begin{equation*}
\begin{aligned}
    (1-\delta)\bar{t}_i(+1)+\delta \bar{t}_i(-1) &\leq \left(\frac{1-2\delta}{1-\delta}\right) \bar{t}_i(+1)+\delta\left(\frac{b-1}{2}\right)\left(\frac{\delta}{1-\delta}\, \bar{f}_i(+1)+\bar{f}_i(-1)\right)\\
    &\leq \left(\frac{b+1-\delta(b+3)}{2}\right)\bar{f}_i(+1)+\left(\frac{\delta(b+3)}{2}-1\right)\bar{f}_i(-1)\\
    &=\left(\frac{b+1}{2}\right)\left((1-\delta)\bar{f}_i(+1)+\delta \bar{f}_i(-1)\right) - \left(\delta \bar{f}_i(+1)+(1-\delta)\bar{f}_i(-1)\right)\,.
\end{aligned}
\end{equation*}
This implies that equation~\eqref{eq:high_type_IR}, which is the high type IIR condition, falls out of the high type BN-IC constraint and the low type IIR constraint.

The above analysis implies that one needs to only maximize the expected transfer on the constrained set induced by the incentive constraints (i.e., equation~\eqref{eq:simplified_BNIC}) and the low type IIR condition. Therefore, at the optimum the low type IIR condition as well as one of the incentive constraints must bind. One can show that since $\delta<1/2$, the extreme point associated with the meet of the low type IIR and high type BN-IC achieves a higher expected revenue than the meet of the low type IIR and low type BN-IC. Hence, the following profile of interim transfers pins down the optimum:
\begin{equation*}
    \begin{aligned}
        \bar{t}_i(-1) &= -\delta \bar{f}_i(+1)+\left(\frac{b-1}{2}+\delta\right)\bar{f}_i(-1)\,,\\
        \bar{t}_i(+1)&= \left(\frac{b+1}{2}-\delta\right)\bar{f}_i(+1)-(1-\delta)\bar{f}_i(-1)\,.
    \end{aligned}
\end{equation*}
Therefore, the maximum expected transfer from individual $i$ is equal to
\begin{equation*}
    \frac{\bar{t}_i(+1)+\bar{t}_i(-1)}{2} = \left(\frac{b+1}{4}-\delta\right)\bar{f}_i(+1)+\left(\frac{b-3}{4}+\delta\right)\bar{f}_i(-1)\,.
\end{equation*}
Since the types are distributed uniformly on $\{-1,+1\}^n$, one has
\begin{equation*}
\begin{aligned}
    \bar{f}_i(+1) &= \BE\left[f\right]+\BE\left[f(x)x_i\right]\,, \\
    \bar{f}_i(-1) &= \BE\left[f\right]-\BE\left[f(x)x_i\right]\,.
\end{aligned}
\end{equation*}
Hence, the maximum expected revenue from implementing $f$ follows:
\begin{equation*}
    \mathsf{R}_\delta[f] = \sum_{i \in [n]}\frac{\bar{t}_i(+1)+\bar{t}_i(-1)}{2} = (1-2\delta) \BE\left[f(x) \sum_{i=1}^n x_i\right]+\left(\frac{b-1}{2}\right)\BE[f(x)]\,,
\end{equation*}
thereby establishing the representation in~\eqref{eq:max_exp_rev}.\qed

\subsection{Proof of Proposition~\ref{prop:surplus_dist}}
The distortion function follows:
\begin{equation*}
\begin{gathered}
    \mathsf{SD}_\delta[f]=\BE\left[\Big(S(x,f(y))- S(x,f(x))\Big)^2\right] = \BE\left[\left(\sum_{i=1}^n\left(\frac{b+x_i}{2}\right)\right)^2 \big(f(x)-f(y)\big)^2 \right]\,.\\
    =\frac{n^2 b^2}{4}\, \BE\left[\big(f(x)-f(y)\big)^2\right] + \frac{nb}{2}\, \BE\left[\left(\sum_{i = 1}^n x_i \right)\big(f(x)-f(y)\big)^2 \right] \\
    +\frac{1}{4}\,\BE\left[\left(\sum_{i = 1}^n x_i \right)^2\big(f(x)-f(y)\big)^2\right]\,.
\end{gathered}
\end{equation*}
Since $f$ is $\{0, 1\}$-valued, it holds that
\begin{equation*}
    \BE\left[\big(f(x)-f(y)\big)^2\right] = \BP\left(f(x) \neq f(y)\right) = \mathsf{NS}_\delta[f]\,.
\end{equation*}
Recall that $\nu_n = \sum_{i=1}^n x_i$. Therefore,
\begin{equation*}
\begin{gathered}
    \left|\frac{1}{n^2}\,\mathsf{SD}_\delta[f] -\frac{b^2}{4}\, \mathsf{NS}_\delta[f] \right| \leq \frac{b}{2}\, \BE\left[\frac{|\nu_n |}{n} \,\big(f(x)-f(y)\big)^2\right] + \BE\left[\frac{\nu_n^2}{4n^2} \, \big(f(x)-f(y)\big)^2\right]\\
    \leq \BE\left[ \frac{b|\nu_n |}{2n} + \frac{\nu_n^2}{4n^2}\right]\,,
\end{gathered}
\end{equation*}
where the second inequality holds because $\big(f(x)-f(y)\big)^2 \leq 1$ for all $f$. Next, observe that $\nu_n/n \to 0$ almost surely, and $|\nu_n /n | \leq 1$ because $x_i\in \{-1,+1\}$. Therefore, by Lebesgue dominated convergence theorem, we have:
\begin{equation*}
    \lim_{n\to \infty} \left|\frac{1}{n^2}\,\mathsf{SD}_\delta[f] -\frac{b^2}{4}\, \mathsf{NS}_\delta[f] \right| \leq \lim_{n\to \infty}  \BE\left[ \frac{b|\nu_n |}{2n} + \frac{\nu_n^2}{4n^2}\right] = 0\,,
\end{equation*}
that concludes the proof of uniform convergence.\qed
\subsection{Proof of Proposition~\ref{prop:maj_rule}}
The expected revenue extracted by the majority rule (followed by equation~\eqref{eq:max_exp_rev}) is equal to
\begin{equation*}
    \mathsf{R}_\delta[f_{\text{maj}}]= (1-2\delta)\BE\left[\nu_n\, \cdot \,\mathbf{1}\left\{\nu_n \geq 0\right\}\right]+\left(\frac{b-1}{2}\right)\BP\left(\nu_n\geq 0\right)\,.
\end{equation*}
Since $\{x_i: i\in [n]\}$ are i.i.d.\, and uniformly distributed $\{-1,+1\}$-valued random variables, then by the central limit theorem $\frac{1}{\sqrt{n}} \nu_n$ converges in distribution to the standard Gaussian, i.e., $Z \sim \mathcal{N}(0,1)$. Therefore, using the Lebesgue dominated convergence theorem one has
\begin{equation*}
    \lim_{n\to \infty}\frac{1}{\sqrt{n}}\,\BE\left[\nu_n^+\right] = \BE\left[Z^+\right]=\frac{1}{\sqrt{2\pi}},\, \text{ and } \, \lim_{n\to \infty}\BP\left(\nu_n\geq 0\right)=\frac{1}{2}\,,
\end{equation*}
thus implying $\mathsf{R}_\delta[f_{\text{maj}}] = \frac{(1-2\delta)}{\sqrt{2\pi}}\sqrt{n}\big(1+o(1)\big)$. 

Next, we examine the noise sensitivity of the majority rule.\footnote{Chapter 5 of \cite{o2014analysis} includes a comprehensive study of the spectral properties of the majority function.} Let $\text{sgn}(\cdot)$ denote the sign function. For the vector of true types $x$, and its noisy variant $y$, one has
\begin{equation*}
    \mathsf{NS}_\delta[f_{\text{maj}}] = \BP\Bigg(\text{sgn}\Big(\sum_{i=1}^n x_i\Big) \neq \text{sgn}\Big(\sum_{i=1}^n y_i\Big)\Bigg)\,,
\end{equation*}
that in turn, due to the symmetry between $x$ and $y$, is equal to twice the following expression
\begin{equation}
\label{eq:NS_majority_rule_expanded}
    \BP\left(\frac{1}{\sqrt{n}} \sum_{i=1}^n x_i \geq 0 \text{ and } \frac{1}{\sqrt{n}} \sum_{i=1}^n y_i<0\right)\,.
\end{equation}
Observe that $\BE\left[x_iy_j\right]=1-2\delta$ when $i=j$ and zero otherwise. Then, because of the multi-dimensional version of central limit theorem, the following weak convergence result holds as $n\to \infty$:
\begin{equation*}
    \left(\frac{1}{\sqrt{n}} \sum_{i=1}^n x_i,\frac{1}{\sqrt{n}} \sum_{i=1}^n y_i\right) \Rightarrow \left(Z_1,\rho Z_1+\sqrt{1-\rho^2}Z_2\right)\,,
\end{equation*}
where $\rho \coloneq 1-2\delta$, and $(Z_1,Z_2)$ are independent standard Gaussians. Hence, the probability in~\eqref{eq:NS_majority_rule_expanded} converges to
\begin{equation*}
    \BP\left(Z_1\geq 0 \text{ and } \rho Z_1+\sqrt{1-\rho^2}Z_2<0\right)\,,
\end{equation*}
which by the rotational symmetry of $(Z_1,Z_2)$ is equal to $\frac{\arccos \rho}{2\pi}$. Therefore,
\begin{equation*}
    \mathsf{NS}_\delta[f_\text{maj}] = \frac{\arccos(1-2\delta)}{\pi} \big(1+o(1)\big)\,.
\end{equation*}\qed
\subsection{Proof of Lemma~\ref{lem:bias_indexing}}
The minimization problem in~\eqref{eq:bias_indexing} clearly falls under the class of linear programs. Therefore, one can express the Lagrangian for this problem as follows:
\begin{equation*}
    \mathcal{L} = \BE\left[f(\nu_n)\right]+\lambda\left(r-\frac{1}{\sqrt{n}}\, \BE\left[f(\nu_n)\nu_n\right]-\frac{b-1}{2(1-2\delta)\sqrt{n}}\, \BE\left[f(\nu_n)\right]\right)\,.
\end{equation*}
The optimal solution thus takes the following form
\begin{equation}
\label{eq:optimal_form}
    f(\nu_n) = \mathbf{1}\left\{\frac{\nu_n}{\sqrt{n}}\geq \frac{1}{\lambda}-\frac{b-1}{2(1-2\delta)\sqrt{n}}\right\}\,.
\end{equation}
Denote the threshold in the above function by $\eta_n\equiv \eta_n(b,\delta,\lambda)$. Since a linear threshold function with the above from is pointwise increasing in $\lambda$, and we want to actually minimize $\BE\left[f(\nu_n)\right]$, then one needs to choose the minimum $\lambda$ that satisfies the revenue constraint, namely:
\begin{equation*}
    \frac{1}{\sqrt{n}}\,\BE\left[f(\nu_n)\nu_n\right]+\frac{b-1}{2(1-2\delta)\sqrt{n}}\, \BE\left[f(\nu_n)\right] \geq r\,.
\end{equation*}
Plugging the optimal form---presented in equation~\eqref{eq:optimal_form}---in the above inequality amounts to:
\begin{equation*}
    \BE\left[\frac{\nu_n}{\sqrt{n}}\, \cdot \, \mathbf{1}\left\{\frac{\nu_n}{\sqrt{n}}\geq \eta_n(b,\delta,\lambda)\right\}\right]+\frac{b-1}{2(1-2\delta)\sqrt{n}}\, \BP\left(\frac{\nu_n}{\sqrt{n}} \geq \eta_n(b,\delta,\lambda)\right) \geq r\,.
\end{equation*}
Applying the central limit theorem followed by monotone convergence theorem imply that as $n \to \infty$, the \textit{lhs} in the above inequality becomes equal to $\varphi(\eta_n)+o(1)$. Therefore, the optimal threshold in equation~\eqref{eq:optimal_form} satisfies:
\begin{equation*}
    \eta_n = \varphi^{-1}(r)+o(1)\,.
\end{equation*}
This verifies the expression for the optimal solution in equation~\eqref{eq:upper_LTF}. Next, one can plug the above finding in equation~\eqref{eq:optimal_form} and obtain an expression for the optimal value of the minimization problem, namely $\alpha_n(r)$:
\begin{equation*}
    \alpha_n(r) = \BP\left(\frac{\nu_n}{\sqrt{n}} \geq \varphi^{-1}(r)+o(1)\right)=\Phi\Big(-\varphi^{-1}(r)\Big)+o(1)\,.
\end{equation*}
The second equality above follows directly from the central limit theorem and thus justifying equation~\eqref{eq:bias_indexing_optimal_value}.\qed
\subsection{The Mapping \texorpdfstring{$\Psi$}.}
\label{app:map}
Let $\Lambda_\rho(x, y)\coloneq \Phi_{\rho}(\Phi^{-1}(x), \Phi^{-1}(y))$. Then, we have $\Psi(x) = x - \Lambda_\rho(x,x)$. We claim that 
\begin{equation}
\label{eq:Lambda_der}
    \frac{\partial \Lambda_\rho(x, y)}{\partial x} = \Phi\left(\frac{\Phi^{-1}(y)-\rho \Phi^{-1}(x)}{\sqrt{1-\rho^2}}\right)\,.
\end{equation}
To see this, let $u=\Phi^{-1}(x)$, and $v = \Phi^{-1}(y)$. Then, 
\begin{equation}
\label{eq:Lambda_chain}
    \frac{\partial \Lambda_\rho(x, y)}{\partial x} = \frac{\partial \Phi_\rho(u, v)}{\partial u} \, \frac{\partial u}{\partial x} = \frac{\partial \Phi_\rho(u, v)}{\partial u} \, \frac{1}{\varphi(u)}\,.
\end{equation}
Let $\varphi_\rho(u, v)$ denote the density associated with $\Phi_\rho(u,v)$. Given $u$, the distribution of $v$ is a Gaussian with mean $\rho u$ and variance $1-\rho^2$, so
\begin{equation*}
    \varphi_\rho(u, v) = \varphi(u) \varphi\left(\frac{v-\rho u}{\sqrt{1-\rho^2}}\right)\,.
\end{equation*}
Therefore, 
\begin{equation*}
    \Phi_\rho(u, v) = \int_{-\infty}^u  \varphi(s) \int_{-\infty}^v  \varphi\left(\frac{t-\rho s}{\sqrt{1-\rho^2}}\right)\d t \, \d s\,,
\end{equation*}
that in turn means
\begin{equation*}
    \frac{\partial \Phi_\rho(u, v)}{\partial u} = \varphi(u)\int_{-\infty}^v \varphi\left(\frac{t-\rho u}{\sqrt{1-\rho^2}}\right)\d t =\varphi(u)  \Phi\left(\frac{v-\rho u}{\sqrt{1-\rho^2}}\right)\,.
\end{equation*}
Substituting the above derivation in~\eqref{eq:Lambda_chain} implies~\eqref{eq:Lambda_der}. Therefore, by the symmetry between $x$ and $y$, it follows that
\begin{equation*}
    \Psi'(x) = 1 - 2\Phi\left(\frac{(1-\rho) \Phi^{-1}(x)}{\sqrt{1-\rho^2}}\right) = 1-2\Phi\left(\sqrt{\frac{1-\rho}{1+\rho}}\, \Phi^{-1}(x)\right)\,.
\end{equation*}
The above function is decreasing in $x$, and is equal to $0$ at $x=1/2$, therefore, $\Psi(x)$ is increasing on $[0,1/2]$ and decreasing on $[1/2, 1]$.
\subsection{Proof Sketch for Proposition~\ref{prop:revenue_equiv_imp_know}}
Following the similar steps of the the proof of Proposition~\ref{prop:revenue_equiv}, we can show that marginal monotonicity is a necessary and sufficient condition for the Bayes-Nash implementability of the SCF $f$. Next, observe that the expected transfer from agent $i$ to the planner is $\big(\bar{t}_i(-1)+\bar{t}_i(+1)\big)/2$. It is then straightforward to show that in the optimum the BN-IC constraint for the high type (namely the first inequality in~\eqref{eq:simplified_BNIC_imp_know}) and the IR condition for the low type (i.e., equation~\eqref{eq:low_type_IR_imp_know}) bind. Hence, the optimum transfers are:
\begin{equation*}
    \begin{aligned}
        \bar{t}_i(-1) &= \left(\frac{b-1}{2}+\delta\right)\bar{f}_i(-1)\,,\\
        \bar{t}_i(+1)&=\left(\frac{b+1}{2}-\delta\right)\bar{f}_i(+1)-(1-2\delta)\bar{f}_i(-1)\,.
    \end{aligned}
\end{equation*}
Given that $\bar{f}_i(z) = \BE[f]+ z\,\BE\left[f(x)x_i\right]$ for $z \in \{-1,+1\}$, summing these expressions over $i$ and then dividing by two results in the representation~\eqref{eq:max_exp_rev_imp_know}.

\section{Intuitive Proof of Lemma~\ref{lem:maj_stablest}}
We present a very high level sketch of the proof, explaining the pillars and the main ideas. The are a handful of different methods for proving this theorem (as recent as~\cite{eldan2022noise}), but we rely on the approach offered in~\cite{Mossel2010noise}.

The proof relies on two main ideas: (\rn{1}) Borell's Gaussian isoperimetric inequality; (\rn{2}) \textit{Invariance principle}. We first present some preliminaries that discipline the reading of how these two ideas come together and shape the proof.
\subsection{Preliminaries}
We start with the definition of the noise operator acting on the Hilbert space $H=L^2\left(\{-1,+1\}^n\right)$ with the uniform measure on the hypercube.
\begin{definition}[Noise operator]
Let $\rho \in (0,1)$ and define $\mathsf{T}_\rho: H\to H$ as
\begin{equation*}
    \mathsf{T}_\rho f (x) = \BE\left[f(y) | x\right]\,,
\end{equation*}
where $x=(x_1,\ldots,x_n)$ is a point uniformly drawn from the hypercube, and $y$ is its $\rho$-correlated version, such that $\BE\left[y_i x_i\right] = \rho$ for each coordinate $i \in [n]$.
\end{definition}

For every basis element $\chi_S \in H$, one has $\mathsf{T}_\rho \chi_S (x) = \rho^{|S|} \chi_S(x)$. Since, the noise operator is linear, applying that on the Fourier expansion in equation~\eqref{eq:Fourier_expansion} implies
\begin{equation*}
    \mathsf{T}_\rho f(x) = \sum_{S \subseteq [n]} \rho^{|S|}\hat{f}(S) \chi_S(x)\,. 
\end{equation*}
In addition, the noise operator is commutative and has the semi-group property, that is for $\rho_1,\rho_2\in (0,1)$, one has $\mathsf{T}_{\rho_1}\mathsf{T}_{\rho_2}=\mathsf{T}_{\rho_2}\mathsf{T}_{\rho_1} = \mathsf{T}_{\rho_1\rho_2}$. Furthermore, the above Fourier representation of the noise operator implies that for every $f,g \in H$, it holds that $\langle f, \mathsf{T}_\rho g \rangle = \langle \mathsf{T}_\rho f ,g\rangle$.

Looking back at the definition of the noise stability in equation~\eqref{eq:noise_stab_def}, one observes that
\begin{equation}
\label{eq:quadratic_exp_for_stability}
    \mathsf{Stab}_\delta [f] = \langle f, \mathsf{T}_{1-2\delta} f\rangle = \langle \mathsf{T}_{\sqrt{1-2\delta}} f ,\mathsf{T}_{\sqrt{1-2\delta}} f \rangle = \BE\left[\left(\mathsf{T}_{\sqrt{1-2\delta}} f(x)\right)^2\right]\,.
\end{equation}

Next, we present the passing from the Boolean to Gaussian environment. Let $\gamma$ be the standard Gaussian measure on $\BR^n$, and $L^2\left(\BR^n;\gamma\right)$ be the Hilbert space of square integrable functions with respect to $\gamma$, equipped with its natural inner product.
\begin{definition}[Gaussian evaluation]
Let $z \in \BR^n$ be distributed according to the standard Gaussian measure $\gamma$. For a Boolean function $f\in H$, we abuse the notation and define its Gaussian evaluation as 
\begin{equation*}
    f(z) = \sum_{S \subseteq [n]} \hat{f}(S) \chi_S(z)\,.
\end{equation*}
Since $f \in H$, then
\begin{equation*}
    \BE_\gamma\left[f(z)^2\right] = \sum_{S \subseteq [n]} \hat{f}(S)^2 = \BE\left[f(x)^2\right]<\infty\,,
\end{equation*}
and hence the Gaussian passing of $f$ belongs to $L^2\left(\BR^n;\gamma\right)$.
\end{definition}
\begin{remark}
Inspired by the previous definition, one can extend the domain of other operators, such $\mathsf{Stab}_\delta$ and $\mathsf{T}_\rho$, to $L^2(\BR^n;\gamma)$. For example, let $z$ and $z'$ be two $n$-dimensional standard Gaussian vectors, where their corresponding coordinates are $\rho$-correlated, then:
\begin{equation*}
\begin{aligned}
    \mathsf{T}_\rho f(z) &= \BE\left[f(z')|z\right] = \sum_{S \subseteq [n]} \rho^{|S|}\hat{f}(S) \chi_S(z)\,,\\
    \mathsf{Stab}_\delta[f] &= \langle f, \mathsf{T}_{1-2\delta} f\rangle = \langle \mathsf{T}_{\sqrt{1-2\delta}} f ,\mathsf{T}_{\sqrt{1-2\delta}} f \rangle = \BE\left[\left(\mathsf{T}_{\sqrt{1-2\delta}} f(z)\right)^2\right]\,.
\end{aligned}
\end{equation*}
\end{remark}

\subsection{Borell's Isoperimetric Inequality}
At this point it is recommended for the reader to refresh their memory with the definitions of Gaussian functions in the remarks~\ref{not:gaussians} and \ref{def:Gaussian_funcs}.
\begin{theorem}[\cite{borell1985geometric}]
Fix $\delta \in (0,1/2)$. Then, for any $f\in L^2(\BR^n;\gamma)$ with the range $[0,1]$, and $\BE[f]=\mu$, it holds that
\begin{equation}
\label{eq:Borell}
    \mathsf{Stab}_\delta [f] \leq \Phi_{1-2\delta}\big(\Phi^{-1}(\mu)\big)\,.
\end{equation}
\end{theorem}
The \textit{rhs} to the above inequality is \textit{equal} to the noise stability of the indicator function of any half-space $H\subseteq \BR^n$ with the Gaussian volume of $\text{Vol}_\gamma(H) = \mu$. 

An interesting corollary to this theorem is that among all measurable subsets of $\BR^n$, with a fixed Gaussian volume, the half-spaces have the minimum sensitivity to noise. Formally, let us denote the $n$-dimensional standard Gaussian probability measure by $\BP_\gamma$. Consider any measurable subset $A$ with $\text{Vol}_\gamma(A) = \mu>0$, and any half-space $H$ with the same volume $\mu$. Then, inequality~\eqref{eq:Borell} implies that
\begin{equation*}
    \BP_\gamma\left(x \in A, y\in A\right) \leq \BP_\gamma\left(x \in H, y\in H\right)\,,
\end{equation*}
where $x \sim \gamma$ and $y$ is its $\delta$-noisy version, that is $\BE[y_i | x_i]= (1-2\delta)x_i$ for each $i\in [n]$. Canceling $\BP_\gamma(x\in A)$ from both sides amounts to
\begin{equation*}
    \BP_\gamma \left(y \in A \big | x\in A\right) \leq \BP_\gamma \left(y \in H \big | x\in H\right)\,.
\end{equation*}
This means if one starts at a random point $x$ inside the subset $A$, then the chances of leaving this region due to adding noise is minimal for half-spaces.
\subsection{Invariance Principle}
In this part, we offer an intuitive statement of the invariance principle. For that, we need to define the concept of \textit{influence}.

Let $x^{i\mapsto +1}$ be the vector $x$, where its $i$-th coordinate is replaced with $+1$. Similarly, define $x^{i\mapsto -1}$. Then, holding all other coordinates constant, one can define the \textit{derivative} operator $\mathsf{D}_i: H\to \BR$ as 
\begin{equation*}
    \mathsf{D}_i[f](x) = \frac{f(x^{i\mapsto +1})-f(x^{i\mapsto -1})}{2}\,.
\end{equation*}
\begin{definition}[Coordinate influence]
For $f:\{-1,+1\}^n \to \BR$ and $i\in [n]$ define
\begin{equation*}
    \mathsf{Inf}_i[f] = \sum_{S \ni i} \hat{f}(S)^2\,.
\end{equation*}
That is the influence of coordinate $i$ on $f$ is the sum of $f$'s squared Fourier weights containing $i$. One can immediately see that $\mathsf{Inf}_i[f] = \BE\left[\mathsf{D}_i[f](x)^2\right]$. Hence, the influence of input $i$ should be interpreted as the expected change that it makes on the function $f$.
\end{definition}
Next, we explain what it means for a function $F:\{-1,+1\}^n \to \BR$ to be \textit{invariant}. For any $x$ not belonging to the hypercube, we identify $F(x)$ by the evaluation of its Fourier representation at $x$. Hence, with some abuse of notation one can extend the domain of $F$ to the entire $\BR^n$. 

Let $x = (x_1,\ldots,x_n)$ and $z = (z_1,\ldots,z_n)$ be two vectors with i.i.d. elements, such that their first few moments match, namely $\BE[x_i] = \BE[x_i^3]=\BE[z_i]=\BE[z_i^3]=0$, $\BE[x_i^2]=\BE[z_i^2]=1$ for all $i\in [n]$, and the fourth moment is finite. For example, $x$ can be drawn uniformly from the hypercube $\{-1,+1\}^n$ and $z$ from the $n$-dimensional standard Gaussian distribution on $\BR^n$. Suppose the previously mentioned function $F$ has \textit{small} influence with respect to all of its input coordinates, that is there is no single coordinate that can determine the outcome with high probability.\footnote{Observe that we intentionally state these results qualitatively, as their quantitative versions require many approximation steps, which are carried out in~\cite{Mossel2010noise}.} Then, the invariance principle claims that for any sufficiently smooth function $\Upsilon:\BR\to \BR$, as $n \to \infty$ one has
\begin{equation}
\label{eq:inv_principle}
    \Big|\BE\left[\Upsilon\big(F(x)\big)\right]-\BE\left[\Upsilon\big(F(z)\big)\right]\Big| = o(1)\,.
\end{equation}
The approximation error $o(1)$ becomes \textit{uniform} over all $F$'s, that put vanishingly small influence on every single coordinate.
\subsection{Proof Sketch}
The reader should now have good senses on how to put the previous two ideas together and reach to the conclusion. First, observe that in our setup, where $f$ is supposed to be an anonymous SCF, the small influence condition automatically holds, because $f$ treats all its input coordinates symmetrically, thus one can safely apply the invariance principle. Second, equation~\eqref{eq:quadratic_exp_for_stability} hints at choosing $\Upsilon$ to be the quadratic function, i.e., $t\mapsto t^2$---that is ``sufficiently smooth''---and to assign $F(x) = \mathsf{T}_{\sqrt{1-2\delta}}f(x)$. Then, the invariance principle in equation~\eqref{eq:inv_principle} implies that
\begin{equation*}
    \Bigg|\BE\left[\left(\mathsf{T}_{\sqrt{1-2\delta}} f(x)\right)^2\right]-\BE\left[\left(\mathsf{T}_{\sqrt{1-2\delta}} f(z)\right)^2\right]\Bigg| = o(1)\,.
\end{equation*}
Third, the Gaussian noise stability is upper bounded by the Borell's isoperimetric inequality in equation~\eqref{eq:Borell}. Hence, the previous equation implies that for every anonymous SCF $f$:
\begin{equation*}
    \mathsf{Stab}_\delta[f] = \BE\left[\left(\mathsf{T}_{\sqrt{1-2\delta}} f(x)\right)^2\right] \leq \Phi_{1-2\delta}\big(\Phi^{-1}(\BE[f])\big)+o(1)\,,
\end{equation*}
thereby verifying the claim of Lemma~\ref{lem:maj_stablest}.

\newpage
\setcitestyle{numbers}	 
\bibliographystyle{ACM-Reference-Format}
\bibliography{ref}


\begin{thebibliography}{34}


\ifx \showCODEN    \undefined \def \showCODEN     #1{\unskip}     \fi
\ifx \showDOI      \undefined \def \showDOI       #1{#1}\fi
\ifx \showISBNx    \undefined \def \showISBNx     #1{\unskip}     \fi
\ifx \showISBNxiii \undefined \def \showISBNxiii  #1{\unskip}     \fi
\ifx \showISSN     \undefined \def \showISSN      #1{\unskip}     \fi
\ifx \showLCCN     \undefined \def \showLCCN      #1{\unskip}     \fi
\ifx \shownote     \undefined \def \shownote      #1{#1}          \fi
\ifx \showarticletitle \undefined \def \showarticletitle #1{#1}   \fi
\ifx \showURL      \undefined \def \showURL       {\relax}        \fi
\providecommand\bibfield[2]{#2}
\providecommand\bibinfo[2]{#2}
\providecommand\natexlab[1]{#1}
\providecommand\showeprint[2][]{arXiv:#2}

\bibitem[Benjamini et~al\mbox{.}(1999)]%
        {benjamini1999noise}
\bibfield{author}{\bibinfo{person}{Itai Benjamini}, \bibinfo{person}{Gil
  Kalai}, {and} \bibinfo{person}{Oded Schramm}.}
  \bibinfo{year}{1999}\natexlab{}.
\newblock \showarticletitle{Noise Sensitivity of Boolean Functions and
  Applications to Percolation}.
\newblock \bibinfo{journal}{\emph{Publications Math{\'e}matiques de l'Institut
  des Hautes {\'E}tudes Scientifiques}}  \bibinfo{volume}{90}
  (\bibinfo{year}{1999}), \bibinfo{pages}{5--43}.
\newblock


\bibitem[Blair et~al\mbox{.}(2015)]%
        {blair2015design}
\bibfield{author}{\bibinfo{person}{Graeme Blair}, \bibinfo{person}{Kosuke
  Imai}, {and} \bibinfo{person}{Yang-Yang Zhou}.}
  \bibinfo{year}{2015}\natexlab{}.
\newblock \showarticletitle{Design and Analysis of the Randomized Response
  Technique}.
\newblock \bibinfo{journal}{\emph{J. Amer. Statist. Assoc.}}
  \bibinfo{volume}{110}, \bibinfo{number}{511} (\bibinfo{year}{2015}),
  \bibinfo{pages}{1304--1319}.
\newblock


\bibitem[Borell(1985)]%
        {borell1985geometric}
\bibfield{author}{\bibinfo{person}{Christer Borell}.}
  \bibinfo{year}{1985}\natexlab{}.
\newblock \showarticletitle{Geometric Bounds on the Ornstein-Uhlenbeck Velocity
  Process}.
\newblock \bibinfo{journal}{\emph{Probability Theory and Related Fields}}
  \bibinfo{volume}{70}, \bibinfo{number}{1} (\bibinfo{year}{1985}),
  \bibinfo{pages}{1--13}.
\newblock


\bibitem[Budish and Kessler(2022)]%
        {budish2022can}
\bibfield{author}{\bibinfo{person}{Eric Budish} {and} \bibinfo{person}{Judd~B.
  Kessler}.} \bibinfo{year}{2022}\natexlab{}.
\newblock \showarticletitle{Can Market Participants Report their Preferences
  Accurately (Enough)?}
\newblock \bibinfo{journal}{\emph{Management Science}} \bibinfo{volume}{68},
  \bibinfo{number}{2} (\bibinfo{year}{2022}), \bibinfo{pages}{1107--1130}.
\newblock


\bibitem[Chen et~al\mbox{.}(2016)]%
        {chen2016truthful}
\bibfield{author}{\bibinfo{person}{Yiling Chen}, \bibinfo{person}{Stephen
  Chong}, \bibinfo{person}{Ian~A. Kash}, \bibinfo{person}{Tal Moran}, {and}
  \bibinfo{person}{Salil Vadhan}.} \bibinfo{year}{2016}\natexlab{}.
\newblock \showarticletitle{Truthful Mechanisms for Agents that Value Privacy}.
\newblock \bibinfo{journal}{\emph{ACM Transactions on Economics and Computation
  (TEAC)}} \bibinfo{volume}{4}, \bibinfo{number}{3} (\bibinfo{year}{2016}),
  \bibinfo{pages}{1--30}.
\newblock


\bibitem[Dwork(2008)]%
        {dwork2008differential}
\bibfield{author}{\bibinfo{person}{Cynthia Dwork}.}
  \bibinfo{year}{2008}\natexlab{}.
\newblock \showarticletitle{Differential Privacy: A Survey of Results}. In
  \bibinfo{booktitle}{\emph{International Conference on Theory and Applications
  of Models of Computation}}. Springer, \bibinfo{pages}{1--19}.
\newblock


\bibitem[Dwork et~al\mbox{.}(2006a)]%
        {dwork2006data}
\bibfield{author}{\bibinfo{person}{Cynthia Dwork}, \bibinfo{person}{Krishnaram
  Kenthapadi}, \bibinfo{person}{Frank McSherry}, \bibinfo{person}{Ilya
  Mironov}, {and} \bibinfo{person}{Moni Naor}.}
  \bibinfo{year}{2006}\natexlab{a}.
\newblock \showarticletitle{Our data, Ourselves: Privacy via Distributed Noise
  Generation}. In \bibinfo{booktitle}{\emph{Annual International Conference on
  the Theory and Applications of Cryptographic Techniques}}. Springer,
  \bibinfo{pages}{486--503}.
\newblock


\bibitem[Dwork et~al\mbox{.}(2006b)]%
        {dwork2006}
\bibfield{author}{\bibinfo{person}{Cynthia Dwork}, \bibinfo{person}{Frank
  McSherry}, \bibinfo{person}{Kobbi Nissim}, {and} \bibinfo{person}{Adam
  Smith}.} \bibinfo{year}{2006}\natexlab{b}.
\newblock \showarticletitle{Calibrating Noise to Sensitivity in Private Data
  Analysis}. In \bibinfo{booktitle}{\emph{Theory of Cryptography}},
  \bibfield{editor}{\bibinfo{person}{Shai Halevi} {and} \bibinfo{person}{Tal
  Rabin}} (Eds.). \bibinfo{publisher}{Springer Berlin Heidelberg},
  \bibinfo{address}{Berlin, Heidelberg}, \bibinfo{pages}{265--284}.
\newblock
\showISBNx{978-3-540-32732-5}


\bibitem[Dwork and Roth(2014)]%
        {dworkroth2014}
\bibfield{author}{\bibinfo{person}{Cynthia Dwork} {and} \bibinfo{person}{Aaron
  Roth}.} \bibinfo{year}{2014}\natexlab{}.
\newblock \showarticletitle{The Algorithmic Foundations of Differential
  Privacy}.
\newblock \bibinfo{journal}{\emph{Foundations and Trends{\textregistered} in
  Theoretical Computer Science}} \bibinfo{volume}{9}, \bibinfo{number}{3--4}
  (\bibinfo{year}{2014}), \bibinfo{pages}{211--407}.
\newblock
\showISSN{1551-305X}
\urldef\tempurl%
\url{https://doi.org/10.1561/0400000042}
\showDOI{\tempurl}


\bibitem[Eilat et~al\mbox{.}(2021)]%
        {eilat2021bayesian}
\bibfield{author}{\bibinfo{person}{Ran Eilat}, \bibinfo{person}{Kfir Eliaz},
  {and} \bibinfo{person}{Xiaosheng Mu}.} \bibinfo{year}{2021}\natexlab{}.
\newblock \showarticletitle{Bayesian Privacy}.
\newblock \bibinfo{journal}{\emph{Theoretical Economics}} \bibinfo{volume}{16},
  \bibinfo{number}{4} (\bibinfo{year}{2021}), \bibinfo{pages}{1557--1603}.
\newblock


\bibitem[Eldan et~al\mbox{.}(2022)]%
        {eldan2022noise}
\bibfield{author}{\bibinfo{person}{Ronen Eldan}, \bibinfo{person}{Dan
  Mikulincer}, {and} \bibinfo{person}{Prasad Raghavendra}.}
  \bibinfo{year}{2022}\natexlab{}.
\newblock \showarticletitle{Noise Stability on the Boolean Hypercube via a
  Renormalized Brownian Motion}.
\newblock \bibinfo{journal}{\emph{arXiv preprint arXiv:2208.06508}}
  (\bibinfo{year}{2022}).
\newblock


\bibitem[Evans and King(2023)]%
        {evans2023statistically}
\bibfield{author}{\bibinfo{person}{Georgina Evans} {and} \bibinfo{person}{Gary
  King}.} \bibinfo{year}{2023}\natexlab{}.
\newblock \showarticletitle{Statistically Valid Inferences from Differentially
  Private Data Releases, with Application to the Facebook Urls Dataset}.
\newblock \bibinfo{journal}{\emph{Political Analysis}} \bibinfo{volume}{31},
  \bibinfo{number}{1} (\bibinfo{year}{2023}), \bibinfo{pages}{1--21}.
\newblock


\bibitem[Evans et~al\mbox{.}(2019)]%
        {evans2019statistically}
\bibfield{author}{\bibinfo{person}{Georgina Evans}, \bibinfo{person}{Gary
  King}, \bibinfo{person}{Margaret Schwenzfeier}, {and}
  \bibinfo{person}{Abhradeep Thakurta}.} \bibinfo{year}{2019}\natexlab{}.
\newblock \showarticletitle{Statistically valid Inferences from Privacy
  Protected Data}.
\newblock \bibinfo{journal}{\emph{American Political Science Review}}
  (\bibinfo{year}{2019}).
\newblock


\bibitem[Evans et~al\mbox{.}(2022)]%
        {evans2022differentially}
\bibfield{author}{\bibinfo{person}{Georgina Evans}, \bibinfo{person}{Gary
  King}, \bibinfo{person}{Adam~D. Smith}, {and} \bibinfo{person}{Abhradeep
  Thakurta}.} \bibinfo{year}{2022}\natexlab{}.
\newblock \showarticletitle{Differentially Private Survey Research}.
\newblock \bibinfo{journal}{\emph{American Journal of Political Science}}
  \bibinfo{volume}{27} (\bibinfo{year}{2022}), \bibinfo{pages}{703--709}.
\newblock


\bibitem[Geng and Viswanath(2015)]%
        {geng2015optimal}
\bibfield{author}{\bibinfo{person}{Quan Geng} {and} \bibinfo{person}{Pramod
  Viswanath}.} \bibinfo{year}{2015}\natexlab{}.
\newblock \showarticletitle{The Optimal Noise-Adding Mechanism in Differential
  Privacy}.
\newblock \bibinfo{journal}{\emph{IEEE Transactions on Information Theory}}
  \bibinfo{volume}{62}, \bibinfo{number}{2} (\bibinfo{year}{2015}),
  \bibinfo{pages}{925--951}.
\newblock


\bibitem[Gleyze and Pernoud(2022)]%
        {gleyze2023}
\bibfield{author}{\bibinfo{person}{Simon Gleyze} {and} \bibinfo{person}{Agathe
  Pernoud}.} \bibinfo{year}{2022}\natexlab{}.
\newblock \bibinfo{title}{How Competition Shapes Information in Auctions}.
  (\bibinfo{year}{2022}).
\newblock
\newblock
\shownote{Mimeo, Stanford University}.


\bibitem[Gymrek et~al\mbox{.}(2013)]%
        {gymrek2013identifying}
\bibfield{author}{\bibinfo{person}{Melissa Gymrek}, \bibinfo{person}{Amy~L.
  McGuire}, \bibinfo{person}{David Golan}, \bibinfo{person}{Eran Halperin},
  {and} \bibinfo{person}{Yaniv Erlich}.} \bibinfo{year}{2013}\natexlab{}.
\newblock \showarticletitle{Identifying Personal Genomes by Surname Inference}.
\newblock \bibinfo{journal}{\emph{Science}} \bibinfo{volume}{339},
  \bibinfo{number}{6117} (\bibinfo{year}{2013}), \bibinfo{pages}{321--324}.
\newblock


\bibitem[Hassidim et~al\mbox{.}(2017)]%
        {hassidim2017mechanism}
\bibfield{author}{\bibinfo{person}{Avinatan Hassidim},
  \bibinfo{person}{D{\'e}borah Marciano}, \bibinfo{person}{Assaf Romm}, {and}
  \bibinfo{person}{Ran~I. Shorrer}.} \bibinfo{year}{2017}\natexlab{}.
\newblock \showarticletitle{The Mechanism is Truthful, Why aren't You?}
\newblock \bibinfo{journal}{\emph{American Economic Review}}
  \bibinfo{volume}{107}, \bibinfo{number}{5} (\bibinfo{year}{2017}),
  \bibinfo{pages}{220--24}.
\newblock


\bibitem[Hassidim et~al\mbox{.}(2021)]%
        {hassidim2021limits}
\bibfield{author}{\bibinfo{person}{Avinatan Hassidim}, \bibinfo{person}{Assaf
  Romm}, {and} \bibinfo{person}{Ran~I. Shorrer}.}
  \bibinfo{year}{2021}\natexlab{}.
\newblock \showarticletitle{The Limits of Incentives in Economic Matching
  Procedures}.
\newblock \bibinfo{journal}{\emph{Management Science}} \bibinfo{volume}{67},
  \bibinfo{number}{2} (\bibinfo{year}{2021}), \bibinfo{pages}{951--963}.
\newblock


\bibitem[He et~al\mbox{.}(2018)]%
        {he2018preserving}
\bibfield{author}{\bibinfo{person}{Jianping He}, \bibinfo{person}{Lin Cai},
  {and} \bibinfo{person}{Xinping Guan}.} \bibinfo{year}{2018}\natexlab{}.
\newblock \showarticletitle{Preserving Data-Privacy with Added Noises: Optimal
  Estimation and Privacy Analysis}.
\newblock \bibinfo{journal}{\emph{IEEE Transactions on Information Theory}}
  \bibinfo{volume}{64}, \bibinfo{number}{8} (\bibinfo{year}{2018}),
  \bibinfo{pages}{5677--5690}.
\newblock


\bibitem[Huang and Kannan(2012)]%
        {huang2012exponential}
\bibfield{author}{\bibinfo{person}{Zhiyi Huang} {and} \bibinfo{person}{Sampath
  Kannan}.} \bibinfo{year}{2012}\natexlab{}.
\newblock \showarticletitle{The Exponential Mechanism for Social Welfare:
  Private, Truthful, and Nearly Optimal}. In \bibinfo{booktitle}{\emph{2012
  IEEE 53rd Annual Symposium on Foundations of Computer Science}}. IEEE,
  \bibinfo{pages}{140--149}.
\newblock


\bibitem[Kalai(2002)]%
        {kalai2002fourier}
\bibfield{author}{\bibinfo{person}{Gil Kalai}.}
  \bibinfo{year}{2002}\natexlab{}.
\newblock \showarticletitle{A Fourier-Theoretic Perspective on the Condorcet
  Paradox and Arrow's Theorem}.
\newblock \bibinfo{journal}{\emph{Advances in Applied Mathematics}}
  \bibinfo{volume}{29}, \bibinfo{number}{3} (\bibinfo{year}{2002}),
  \bibinfo{pages}{412--426}.
\newblock


\bibitem[Kr{\"a}hmer and Strausz(2023)]%
        {krahmer2023optimal}
\bibfield{author}{\bibinfo{person}{Daniel Kr{\"a}hmer} {and}
  \bibinfo{person}{Roland Strausz}.} \bibinfo{year}{2023}\natexlab{}.
\newblock \showarticletitle{Optimal Nonlinear Pricing with Data-Sensitive
  Consumers}.
\newblock \bibinfo{journal}{\emph{American Economic Journal: Microeconomics}}
  \bibinfo{volume}{15}, \bibinfo{number}{2} (\bibinfo{year}{2023}),
  \bibinfo{pages}{80--108}.
\newblock


\bibitem[McFadden(2009)]%
        {mcfadden2009}
\bibfield{author}{\bibinfo{person}{Daniel McFadden}.}
  \bibinfo{year}{2009}\natexlab{}.
\newblock \showarticletitle{The Human Side of Mechanism Design: a Tribute to
  Leo Hurwicz and Jean-Jacque Laffont}.
\newblock \bibinfo{journal}{\emph{Review of Economic Design}}
  \bibinfo{volume}{13}, \bibinfo{number}{1} (\bibinfo{year}{2009}),
  \bibinfo{pages}{77--100}.
\newblock


\bibitem[McSherry and Talwar(2007)]%
        {mcsherry2007mechanism}
\bibfield{author}{\bibinfo{person}{Frank McSherry} {and} \bibinfo{person}{Kunal
  Talwar}.} \bibinfo{year}{2007}\natexlab{}.
\newblock \showarticletitle{Mechanism Design via Differential Privacy}. In
  \bibinfo{booktitle}{\emph{48th Annual IEEE Symposium on Foundations of
  Computer Science (FOCS'07)}}. IEEE, \bibinfo{pages}{94--103}.
\newblock


\bibitem[Mossel et~al\mbox{.}(2010)]%
        {Mossel2010noise}
\bibfield{author}{\bibinfo{person}{Elchanan Mossel}, \bibinfo{person}{Ryan
  O'Donnell}, {and} \bibinfo{person}{Krzysztof Oleszkiewicz}.}
  \bibinfo{year}{2010}\natexlab{}.
\newblock \showarticletitle{Noise Stability of Functions with Low Influences:
  Invariance and Optimality}.
\newblock \bibinfo{journal}{\emph{Annals of Mathematics}}
  \bibinfo{volume}{171} (\bibinfo{year}{2010}), \bibinfo{pages}{295--341}.
\newblock


\bibitem[Nissim et~al\mbox{.}(2012)]%
        {nissim2012approximately}
\bibfield{author}{\bibinfo{person}{Kobbi Nissim}, \bibinfo{person}{Rann
  Smorodinsky}, {and} \bibinfo{person}{Moshe Tennenholtz}.}
  \bibinfo{year}{2012}\natexlab{}.
\newblock \showarticletitle{Approximately Optimal Mechanism Design via
  Differential Privacy}. In \bibinfo{booktitle}{\emph{Proceedings of the 3rd
  Innovations in Theoretical Computer Science conference}}
  \emph{(\bibinfo{series}{ITCS '12})}. \bibinfo{publisher}{Association for
  Computing Machinery}, \bibinfo{address}{New York, NY, USA},
  \bibinfo{pages}{203--213}.
\newblock


\bibitem[Nissim and Xiao(2015)]%
        {nissim536761}
\bibfield{author}{\bibinfo{person}{Kobbi Nissim} {and} \bibinfo{person}{David
  Xiao}.} \bibinfo{year}{2015}\natexlab{}.
\newblock \bibinfo{booktitle}{\emph{Mechanism Design and Differential
  Privacy}}.
\newblock \bibinfo{publisher}{Springer Berlin Heidelberg},
  \bibinfo{address}{New York, NY}. 1--12 pages.
\newblock
\urldef\tempurl%
\url{http://link.springer.com/referenceworkentry/10.1007/978-3-642-27848-8_548-1}
\showURL{%
\tempurl}


\bibitem[O'Donnell(2014)]%
        {o2014analysis}
\bibfield{author}{\bibinfo{person}{Ryan O'Donnell}.}
  \bibinfo{year}{2014}\natexlab{}.
\newblock \bibinfo{booktitle}{\emph{Analysis of Boolean Functions}}.
\newblock \bibinfo{publisher}{Cambridge University Press}.
\newblock


\bibitem[Rees-Jones(2018)]%
        {rees2018suboptimal}
\bibfield{author}{\bibinfo{person}{Alex Rees-Jones}.}
  \bibinfo{year}{2018}\natexlab{}.
\newblock \showarticletitle{Suboptimal Behavior in Strategy-Proof Mechanisms:
  Evidence from the Residency Match}.
\newblock \bibinfo{journal}{\emph{Games and Economic Behavior}}
  \bibinfo{volume}{108} (\bibinfo{year}{2018}), \bibinfo{pages}{317--330}.
\newblock


\bibitem[Rees-Jones and Skowronek(2018)]%
        {rees2018experimental}
\bibfield{author}{\bibinfo{person}{Alex Rees-Jones} {and}
  \bibinfo{person}{Samuel Skowronek}.} \bibinfo{year}{2018}\natexlab{}.
\newblock \showarticletitle{An Experimental Investigation of Preference
  Misrepresentation in the Residency Match}.
\newblock \bibinfo{journal}{\emph{Proceedings of the National Academy of
  Sciences}} \bibinfo{volume}{115}, \bibinfo{number}{45}
  (\bibinfo{year}{2018}), \bibinfo{pages}{11471--11476}.
\newblock


\bibitem[Thereze(2022)]%
        {thereze2023}
\bibfield{author}{\bibinfo{person}{{Jo\~{a}o} Thereze}.}
  \bibinfo{year}{2022}\natexlab{}.
\newblock \bibinfo{title}{Adverse Selection and Endogenous Information}.
  (\bibinfo{year}{2022}).
\newblock
\newblock
\shownote{Mimeo, Princeton University}.


\bibitem[Warner(1965)]%
        {warner1965randomized}
\bibfield{author}{\bibinfo{person}{Stanley~L. Warner}.}
  \bibinfo{year}{1965}\natexlab{}.
\newblock \showarticletitle{Randomized Response: A Survey Technique for
  Eliminating Evasive Answer Bias}.
\newblock \bibinfo{journal}{\emph{J. Amer. Statist. Assoc.}}
  \bibinfo{volume}{60}, \bibinfo{number}{309} (\bibinfo{year}{1965}),
  \bibinfo{pages}{63--69}.
\newblock


\bibitem[Xiao(2013)]%
        {xiao2013privacy}
\bibfield{author}{\bibinfo{person}{David Xiao}.}
  \bibinfo{year}{2013}\natexlab{}.
\newblock \showarticletitle{Is Privacy Compatible with Truthfulness?}. In
  \bibinfo{booktitle}{\emph{Proceedings of the 4th Conference on Innovations in
  Theoretical Computer Science}} \emph{(\bibinfo{series}{ITCS '13})}.
  \bibinfo{publisher}{Association for Computing Machinery},
  \bibinfo{pages}{67--86}.
\newblock


\end{thebibliography}
\end{document}